\documentclass[12pt,letterpaper]{article}
\usepackage[margin=1.5in]{geometry}
\usepackage[english]{babel}
\usepackage[utf8x]{inputenc}
\usepackage{amsmath}
\usepackage{amssymb} 
\usepackage[retainorgcmds]{IEEEtrantools}
\usepackage{graphicx}
\usepackage{tabularx}
\usepackage{subfig}
\usepackage{kpfonts}    
\usepackage{microtype} 
\usepackage{booktabs}   
\usepackage{bm}         
\usepackage{listings}   
\usepackage{verbatim}   
\usepackage{color}  
\usepackage[colorlinks=true, citecolor=blue]{hyperref}
\usepackage[colorinlistoftodos]{todonotes}
\usepackage{natbib}
\usepackage{setspace}
\usepackage{booktabs}       
\usepackage{amssymb}        
\usepackage{threeparttable} 
\usepackage{caption}

\begin{document}
\title{\textbf{Beyond Patents: R\&D, Capital, and the Productivity Puzzle in Early-Stage High-Tech Firms}}
\author{Victor (Xucheng) CHEN}
\date{7/21/2025}
\maketitle

\begin{abstract}
This study investigates the relationship between innovation activities and firm-level productivity among early-stage high-tech startups in China. Using a proprietary dataset encompassing patent records, R\&D expenditures, capital valuation, and firm performance from 2020 to 2024, we examine whether and how innovation—measured by patents and R\&D input—translates into economic output. Contrary to established literature, we find that patent output does not significantly contribute to either income or profit among the sampled firms. Further investigation reveals that patents may primarily serve a signaling function to external investors and policymakers, rather than reflecting true innovative productivity. In contrast, R\&D expenditure shows a consistent and positive association with firm performance. Through mechanism analysis, we explore three channels—organizational environment, employee quality, and policy-driven incentives—to explain the impact of R\&D, identifying capital inflow and valuation as key drivers of R\&D investment. Finally, heterogeneity analysis indicates that the effects of R\&D are more pronounced in sub-industries such as smart terminals and digital creativity, and for firms based in Shenzhen. Our findings challenge the prevailing assumption that patent output is a universal indicator of innovation success and underscore the context-dependent nature of innovation-performance linkages in emerging markets.
\end{abstract}

\newpage
\section{Introduction and Literature Review}

\subsection{Innovation Has Links with Productivity, and Most Papers Reveal They Are Positively Correlated Using Patents as a Proxy}
\label{subsec:mature}

The evaluation of productivity is becoming increasingly important—not only for providing accurate representations of economies, but also for offering practical insights into how productivity can be enhanced.

Productivity itself can be measured through different indicators, including market value, profitability, and revenue growth \citep{aipatentfail}. Existing studies have largely confirmed a positive relationship between innovation and productivity, suggesting that firms with stronger innovative capacities tend to perform better.For instance, \citet{acs1990innovation} compare static and dynamic measures of small-firm viability and find a positive correlation between innovation and firm performance. However, they also emphasize that the interaction between innovation and productivity is complex and context-dependent, warranting further examination. \citet{geroski1993profitability} propose that innovation may have both direct (but temporary) effects on profitability—via the introduction of new products or processes—and indirect effects, as it signals internal capability shifts within firms. In addition, innovation is found to mediate firm market value \citep{https://doi.org/10.1002/smj.461}, and internally generated knowledge is positively associated with superior firm outcomes \citep{CARDINAL2000247}.

These studies highlight the multifaceted and non-linear nature of the innovation–productivity nexus. Yet, most existing literature draws from mature firms in well-established economies. This raises the question of whether such findings generalize to firms at early stages of development, or to specific industries where innovation may take different forms.

Against this backdrop, patents have emerged as a widely accepted proxy for measuring innovation. Numerous studies have validated the high positive correlation between patent activity and innovation capability \citep{patentindex1,patentindex2}, making patent data a practical input for empirical estimation. \citet{https://doi.org/10.1111/joie.12050}, for instance, directly adopt patents as a proxy for both innovation and productivity. Building on this, \citet{griliches1991r} provide one of the foundational discussions on using patent data to capture R\&D output and firm-level technological progress.

Furthermore, a growing body of research leverages patent statistics to study innovation-driven firm performance. For example, using regression analysis on over 30,000 patent records from 42 firms, \citet{keybymyself1} show that a firm’s growth is positively associated with its capacity to generate rare and valuable knowledge, and to build upon that knowledge. In competitive markets, patent-based metrics reveal accelerated technological progress \citep{patentindex4}. Other scholars observe that patenting activity correlates with higher academic productivity among university inventors \citep{patentindex5}, reinforcing the value of patents as an indicator of innovation output.

While the above studies establish the general link between patent-based innovation and firm productivity, they tend to focus on publicly listed firms in developed markets, where data is more accessible. Consequently, the understanding of this relationship in \textbf{early-stage firms}, especially in \textbf{emerging economies like China}, remains limited. Moreover, few studies attempt to \textbf{disaggregate findings by industry}—even though innovation intensity and pathways may vary greatly between, say, the pharmaceutical sector and the semiconductor sector. These gaps form the entry point for this study, which seeks to explore the patent–productivity relationship in China's early-stage high-tech firms using firm-level panel data.

\subsection{Pure Benefits of Patents to Productivity? Maybe Not.}

Although numerous studies suggest a positive correlation—and even causality—between innovation and productivity, there are two important caveats worth emphasizing.

First, patents are widely used as proxies for innovation, yet they do not always contribute positively to productivity. In some cases, patents may even have a negative or negligible effect on firm performance, while other aspects of innovation, such as R\&D efforts or organizational learning, still exhibit a positive relationship with productivity \citep{acs1990innovation}. This raises a crucial question: to what extent do patents themselves, independent of other innovation inputs, drive productivity growth? This question underscores the relevance of our study, which focuses not on the broad concept of innovation, but on isolating the specific contribution of patents to firm productivity.

Second, a long-standing concern in the literature is the so-called "patent-productivity paradox"—the observation that productivity growth has often lagged behind the surge in patenting activities. This suggests that even when patents do reflect innovation, they do not necessarily translate into improved firm performance. This paradox highlights the need to examine not just whether patents proxy for innovation, but whether they actually contribute to productivity gains. Several studies have sought to explain this disconnect between increases in total factor productivity (TFP) and patent counts. 

Some researchers argue that patents have become nominal or strategic indicators rather than reflections of true innovative output. For instance, \citet{SWEET201978} find that patent rights have no measurable impact on productivity growth. In the context of AI-related industries, where patents are used as proxies for technological advancement, scholars have shown that AI patenting remains a niche phenomenon with limited diffusion and influence \citep{aipatentfail}. Other studies have even documented a negative relationship between patenting and firm performance indicators such as return on assets (ROA) and sales growth \citep{https://doi.org/10.1111/j.1540-5885.2010.00747.x}.

Several potential mechanisms have been proposed to explain these patterns. One explanation centers on the rise of strategic patenting: firms increasingly use patents as competitive tools rather than genuine innovation outputs, leading to a proliferation of low-impact or redundant patents that do not enhance firm productivity \citep{https://doi.org/10.1111/j.1540-5885.2010.00747.x}. Another explanation questions the validity of patents as innovation proxies. If firms are primarily generating incremental rather than breakthrough patents, then the correlation between patent counts and productivity is weakened. \citet{ahuja2001entrepreneurship} identify three organizational pathologies that inhibit radical innovation—the familiarity trap, the maturity trap, and the propinquity trap—which tend to afflict mature firms more than young ventures. Lastly, \citet{cockburn2023canada} argue that the weakening link between patents and productivity in contexts such as Canada may be driven not by invention quality or sectoral bias, but by structural issues such as foreign patent ownership and inventor migration.

Taken together, these contrasting findings suggest that the productivity effects of patents are ambiguous and highly context-dependent, varying across industries, stages of firm development, and national settings.

However, many of the mechanisms proposed in the literature are based on observations of mature firms in developed economies, and may not be applicable to the early-stage, high-tech start-ups that operate under different institutional and organizational conditions. We examine this contextual mismatch in greater detail in the next section. Despite extensive research, we still lack clear answers to a critical question: what is the relationship between patenting and productivity in \textbf{early-stage, high-tech firms in emerging economies like China}? This gap in the literature calls for more targeted, firm-level empirical studies focusing on high-technology start-ups—firms that may exhibit unique innovation patterns and commercialization constraints that are not captured by studies of large or publicly listed companies.

\subsection{The Need to Understand Early-stage Dynamics \& Gaps in Emerging Chinese High-tech Economy Contexts}
\label{sub: our context}

Firstly, there is a notable lack of research focusing on unlisted early-stage firms, especially in Chinese high-tech industries. Most existing studies rely on financial databases covering publicly listed companies that have already undergone IPOs and reached a more mature stage. (See \citet{ernst2001patent} for a discussion of existing empirical work in this area. See also \ref{subsec:mature} for more discussion.) However, early-stage firms operate in entirely different institutional, financing, and commercialization environments. (See \citet{schneider2010young} for a discussion of the relevant policy issues.) This divergence calls for focused investigation: what dynamics shape the relationship between patenting and productivity in nascent, rapidly evolving firms? Besides, many studies implicitly assume that innovation—particularly in the form of patents—is implemented efficiently and yields direct productivity gains. Yet early-stage firms are uniquely agile and vulnerable. They may engage in patenting activities without clear strategies for monetization or fail to translate patents into scalable operational processes. Consequently, assuming a universally positive productivity benefit from patenting may be misleading. Furthermore, most existing studies systematically select firms with already-successful patent outcomes \citep{UKstarups}, thus overlooking the broader population of early-stage innovators. This selection bias presents an opportunity to study the unobserved tail of less successful or still-developing firms.

Secondly, the commonly cited patent inefficiencies and explanations for mature firms—such as strategic patenting \citep{https://doi.org/10.1111/j.1540-5885.2010.00747.x} or path dependency—may \citep{ahuja2001entrepreneurship} not readily apply to early-stage start-ups. These younger firms, especially in emerging sectors, are less constrained by legacy systems and often retain full ownership over their intellectual property. Therefore, understanding their innovation–productivity dynamics requires new conceptual and empirical lenses that are not reducible to those used for large incumbents.

Thirdly, while a few studies have explored start-up dynamics, they often fall short of answering our core questions. For instance, \citet{USsoftwarestartups} manually categorize software start-ups into three stages and argue that for pre-revenue firms, patents offer little value—investors focus instead on market potential and managerial competence. In contrast, for revenue-generating firms, patents play a role in signaling sustainable differentiation. While insightful, their framework emphasizes \textbf{patents as a signal to investors}, rather than considering \textbf{patents as a driver of firm productivity}. This framing introduces serious endogeneity: if venture capital (VC) investment is both a consequence and cause of patenting, it becomes difficult to isolate the true effect of patents on firm performance. In reality, these two effects may reinforce each other, but disentangling them is vital. In our context—Chinese high-tech start-ups—the intrinsic innovation value of patents may outweigh their signaling value. We therefore argue for the importance of controlling for investment status, funding needs, and evaluation-based expectations when assessing patent effectiveness. This approach helps us move beyond correlation and toward causal understanding. Moreover, their work does not account for early-stage firms that may no longer seek VC funding, such as those already operating profitably after angel investment. Nor do they address the productivity role of patents after funding has been secured. These conceptual and methodological gaps give us space to develop new contributions—particularly by controlling for funding stages and by modeling different startup growth paths.

Fourthly, we must recognize the distinctiveness of high-tech industries. These sectors remain “blue oceans” in China, marked by rapid innovation, fierce competition, and uncertain technological trajectories \citep{chen2005high}. In such environments, patents may function differently—sometimes as strategic signals, other times as innovation enablers. These contexts demand empirical evidence that is both granular and context-sensitive.

In summary, we aim to shed light on the role of patenting in small, early-stage, high-tech Chinese firms—a group that remains empirically under-explored, largely due to data constraints. Even among the few papers that focus on start-ups, such as \citet{USsoftwarestartups}, the role of patents as innovation instruments rather than investor signals remains understudied. Our study helps fill this gap.

\subsection{Our Contribution}

This paper contributes to the literature by leveraging a unique panel dataset of early-stage, high-tech firms in China, containing annual firm-level observations on revenue and patenting activity. We investigate whether patenting behavior is correlated with firm productivity, proxied by revenue changes, and whether this relationship varies across firm sizes, sectors, and stages of development. Our empirical design explicitly accounts for potential endogeneity arising from investment status, allowing us to disentangle the signaling versus intrinsic innovation value of patents. This study offers new insights into the innovation–performance nexus for nascent firms and provides policy-relevant evidence on whether patent-based innovation support leads to meaningful economic outcomes.3

\section{Data and Variables}
\label{data part}
This study utilizes sensitive and unpublished data obtained from an anonymous VC fund that consistently invests in emerging high-tech startups. Each time a sub-fund of this VC invests in an early-stage company, it follows up with detailed operational tracking to assess firm performance and prepare intelligence reports for subsequent funding rounds. These reports form the basis of our data source and are highly valuable for research purposes.

The uniqueness and sensitivity of this dataset enable us to investigate early-stage innovation activities using panel data spanning multiple years—a rare opportunity, as most existing studies rely on publicly listed firms that have already succeeded in transforming innovation into productivity. This fills a major empirical gap in the literature. Moreover, the panel structure across several years allows us to incorporate firm and year fixed effects, improving causal inference. Since the data is collected not through self-reporting but as part of internal due diligence for external investors, it is of high accuracy and largely free from self-reporting bias.

\begin{figure}[htbp]
    \centering
    \includegraphics[width=0.3\textwidth]{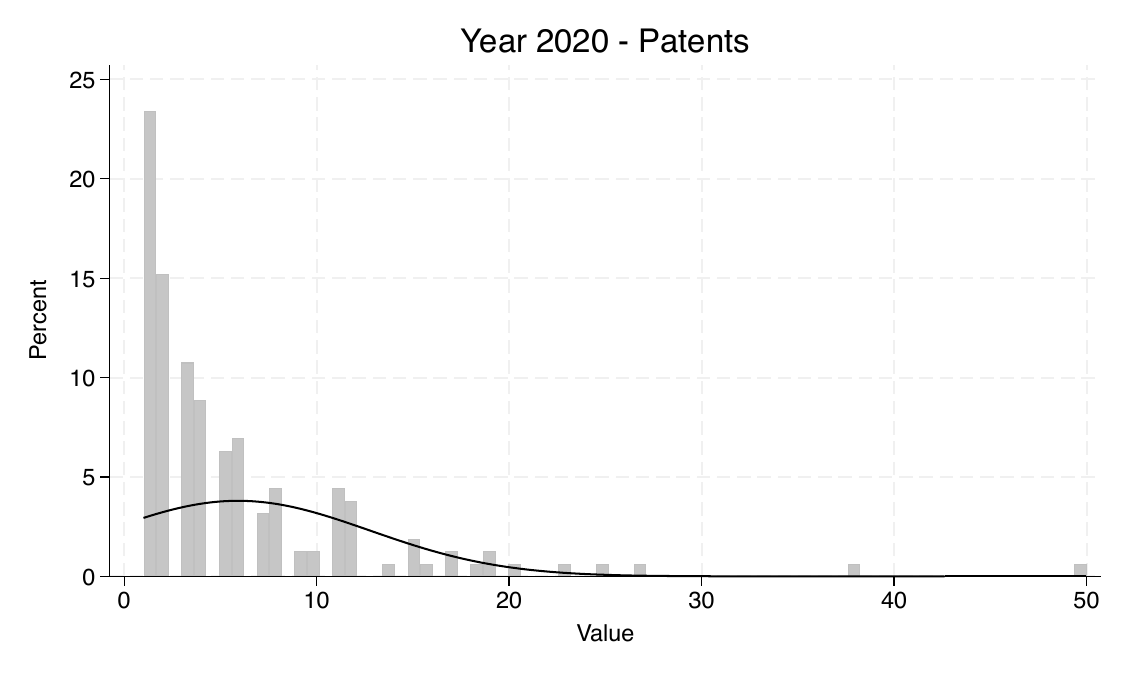}
    \includegraphics[width=0.3\textwidth]{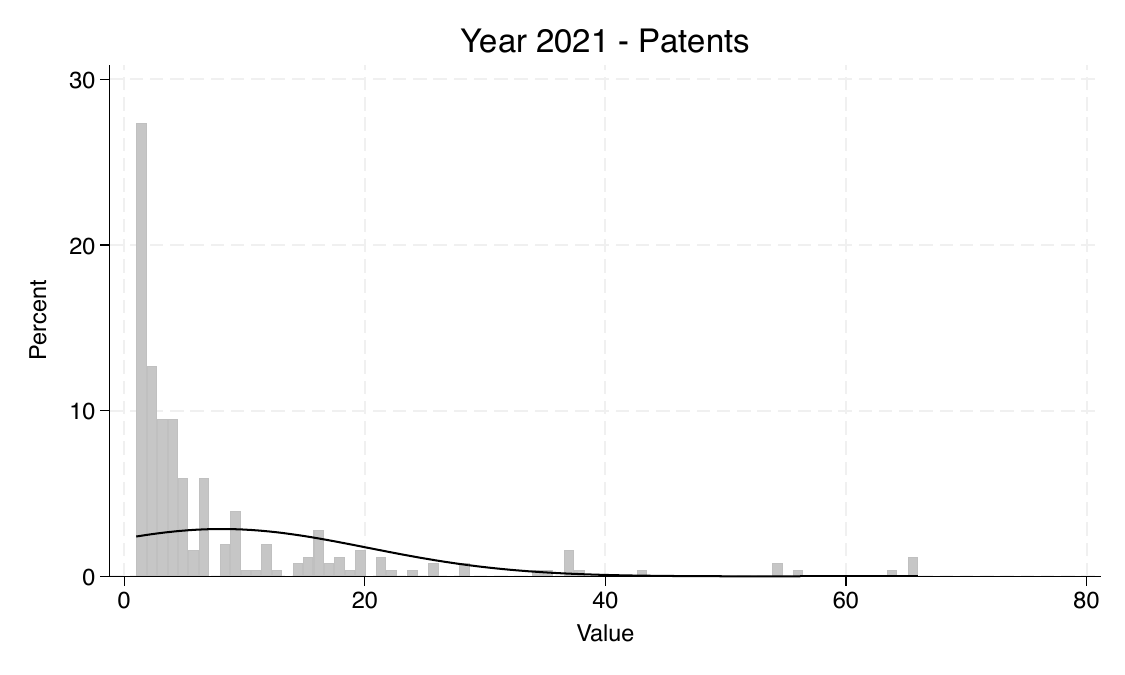}
    \includegraphics[width=0.3\textwidth]{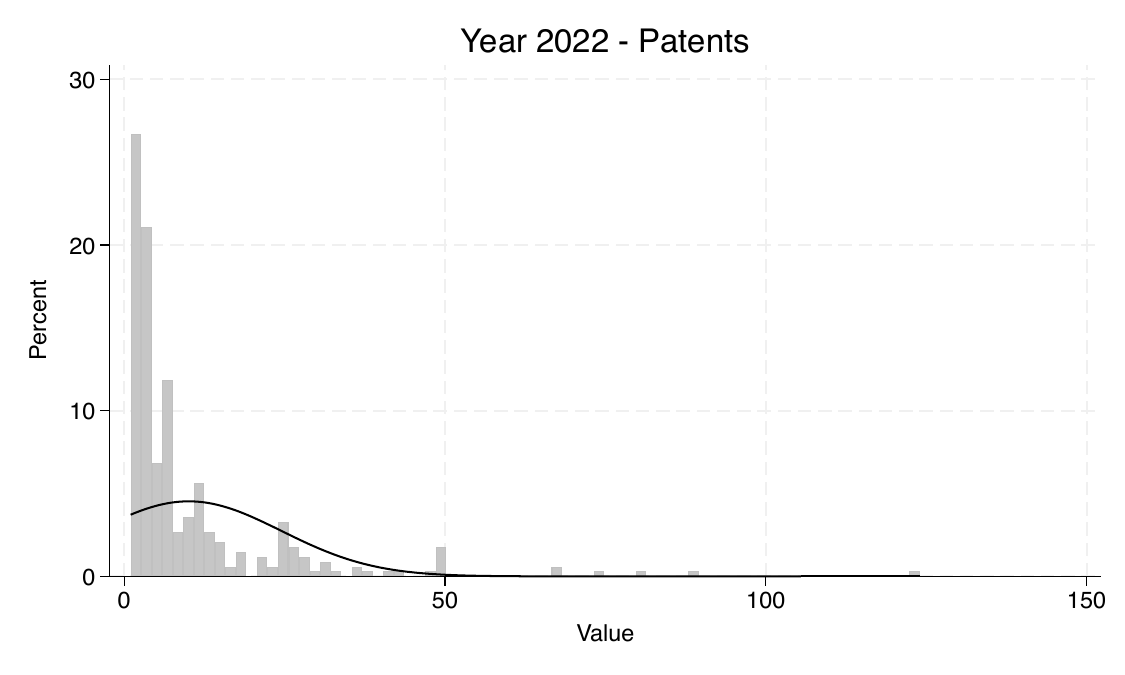}
    
    \vspace{0.2cm}
    \includegraphics[width=0.3\textwidth]{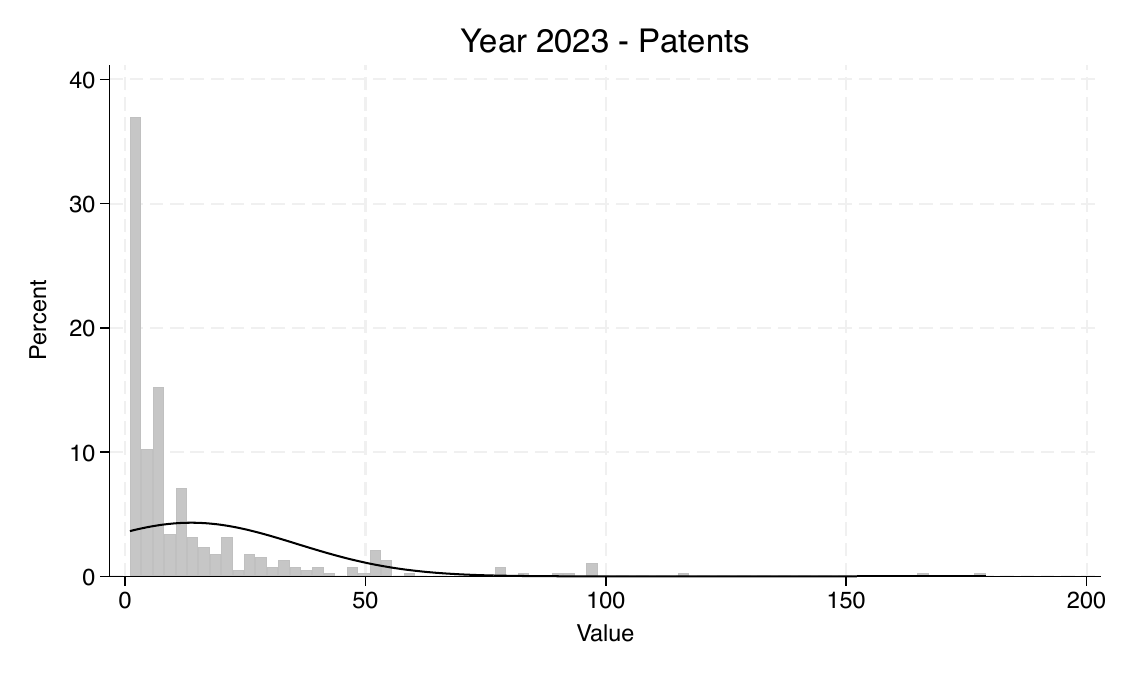}
    \includegraphics[width=0.3\textwidth]{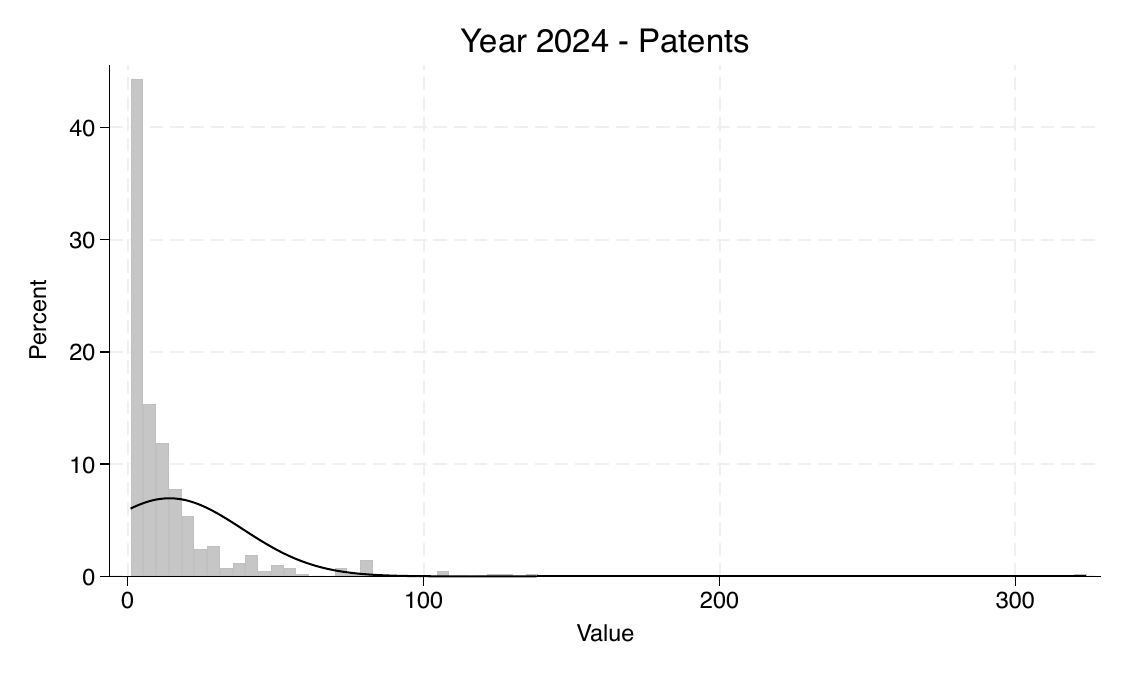}

    \caption{Distribution of Patents (2020–2024)}
    \label{fig:patents_years}
\end{figure}

The dataset contains information on 628 high-tech startups across 88 distinct variables, including annual net profits, R\&D investments, total operating costs, revenues, and company names from 2020 to 2024. Based on firm identifiers, we collected patent data from Wind—a public investment information platform that synchronizes with China’s official patent registration system (see Figure \ref{fig:patents_years}). This yields a rich panel dataset. Furthermore, the data documents whether each firm has received government designations such as “National High-Tech Enterprise,” “Innovative SME,” or “Specialized and Innovative SME.” The significance of these titles will be discussed in the empirical section.

However, the dataset's advantage—its detailed and accurate records for outside investors—also leads to a limitation: some variables are only available for 2024. For example, firm valuation, number of employees, development stage (as classified by Chinese VC practices), and most recent funding round are all recorded only for that year. Despite this limitation, we use the 2024 cross-sectional data for robustness checks and channel analyses, focusing on variables that are relatively time-invariant and structurally representative across firms.

Another important point worth emphasizing is that the start-ups in our dataset were registered at different times and in different years (see Table \ref{tab:regyear} for more details), which is both expected and meaningful. This temporal dispersion reinforces our focus on early-stage companies and minimizes concerns about selection bias. Some firms have shown promising growth, while others exhibit substantial risk factors as documented in their respective reports. Similarly, certain companies have filed patents, whereas others have not, for a variety of reasons.

Upon reviewing the annual reports obtained from investment information platforms such as Wind and corporate research tools like Qichacha, we find that these companies generally lack homogeneous traits or shared characteristics. This observation is consistent with the operational model of the venture capital firm that originally selected them. When the VC firm invested in these companies, it had no precise foresight into how each would perform in the coming years. In this sense, the selection process appears quasi-random, which enhances the representativeness of our sample within the broader context of China’s high-tech start-up landscape.
\begin{table}[htbp]
\centering
\caption{Distribution of Firms by Registration Year}
\label{tab:regyear}
\begin{tabular}{lrr}
\toprule
\textbf{Registration Year} & \textbf{Number} & \textbf{Percentage (\%)} \\
\midrule
2000 & 1   & 0.2  \\
2003 & 1   & 0.2  \\
2010 & 1   & 0.2  \\
2011 & 1   & 0.2  \\
2012 & 1   & 0.2  \\
2014 & 5   & 0.8  \\
2015 & 8   & 1.3  \\
2016 & 23  & 3.7  \\
2017 & 56  & 8.9  \\
2018 & 98  & 15.6 \\
2019 & 83  & 13.2 \\
2020 & 110 & 17.5 \\
2021 & 129 & 20.5 \\
2022 & 56  & 8.9  \\
2023 & 37  & 5.9  \\
2024 & 16  & 2.5  \\
2025 & 2   & 0.3  \\
\midrule
\textbf{Total} & \textbf{628} & \textbf{100.0} \\
\bottomrule
\end{tabular}
\end{table}

We also want to highlight several important facts and hypotheses regarding the companies in our dataset. Due to data limitations, many of these hypotheses are difficult to verify empirically, but we retain them for theoretical reasoning. First, we argue that for the start-ups in our context, time-invariant factors are relatively stable and exert limited influence on our results. Although we incorporate fixed effects in our main regressions, it is still useful to understand this fact—particularly for validating our channel testing or robustness checks when panel fixed effects are not feasible.

Take, for example, the characteristics of the founder. Founders of high-tech start-ups are highly likely to share similar traits—being ambitious, well-educated, and knowledgeable about the industry \citep{wu2009competency}. We argue that it is unlikely for a founder’s managerial style or the broader organizational environment to experience dramatic changes within a short period. In addition, \citet{swiercz2002entrepreneurial} finds that founders in high-tech start-ups have limited impact on firm productivity. These findings give us confidence to conduct robustness checks and cross-sectional analyses using the 2024 data as a meaningful aggregation of the prior four years.

\begin{figure}[htbp]
    \centering
    \includegraphics[width=0.75\textwidth]{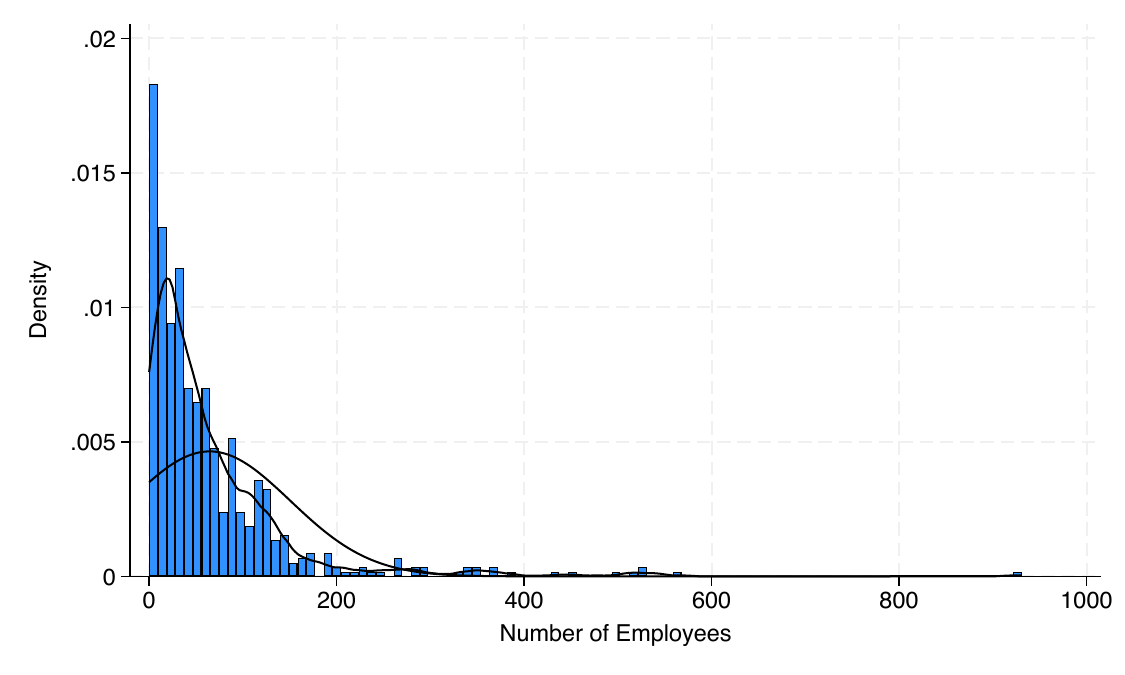}
    \caption{Histogram of Number of Employees (Density)}
    \label{fig:hist_employees}
\end{figure}

Another noteworthy concern involves confounding variables, particularly firm size (see Table \ref{fig:hist_employees}). Due to data constraints, we lack employee headcounts for each year, as most annual reports did not disclose this information. Nonetheless, we provide several justifications for why our available firm size data (collected in 2024) are sufficient. First, based on company recruitment plans available in our dataset, the majority of firms already have enough employees to handle their current operations. Most do not plan to expand their workforce. Approximately 70 companies (about 10\% of the dataset) have posted recruitment needs, and among them, around 68\% are hiring solely to support business expansion. Importantly, these companies report having maintained a relatively stable team size prior to 2024.

Furthermore, according to records from their VC funding rounds, most firms hired key employees during their first year of registration—an expected move for start-ups trying to minimize fixed costs. Once established, team compositions tend to remain stable. Public sources and business intelligence platforms show that most firms had fewer than 10 changes in personnel over the years, which is reasonable given their small initial scale. This observation is further supported by \citet{tai2025effects}, who argues that firms with stable environments and business models are less likely to experience large fluctuations in headcount.

Additionally, prior research suggests that product improvement diminishes the positive effect of firm size on new product innovation \citep{acs1996innovation}. Given that most of our sampled firms focus on improving existing technologies rather than inventing entirely new products, this further reduces the relevance of firm size as a confounding factor. Therefore, we are confident that potential biases related to firm size are minimal. Where necessary, we control for firm size through standardization—for instance, by taking logarithmic transformations or dividing variables by the number of employees—thus mitigating residual distortion to an acceptable degree.

Another potential concern we aim to address—closely aligned with the prior literature—is the signaling effect of patents in start-ups, as discussed by \citet{UKstarups}. Specifically, we seek to clarify why our dataset holds unique empirical value in this context. Unlike prior studies, where firms may utilize patents developed prior to their official founding or before receiving venture capital (VC) support, our dataset strictly confines patent activity to the period \textit{after} the first VC investment and \textit{before} the most recent external investment as of 2024. This time window allows us to focus on innovation occurring within a well-defined and comparable stage of company development, while ensuring that all firms have already secured their initial funding. By restricting our patent data to the post-VC operational period, we effectively reduce the confounding effect of patents serving as ex ante investment signals and instead focus on their genuine contribution to innovation and productivity.

However, it is important to note that this approach cannot entirely eliminate the possibility that patents still function, to some extent, as a signaling mechanism. Some firms continue to seek external investment after their first VC round, making it difficult to fully disentangle whether their early-stage patenting activities are driven by the intention to achieve technological innovation or to attract further investment. Nevertheless, we argue that this restriction significantly mitigates the signaling bias, allowing us to better identify the productivity-enhancing role of patents in the post-VC phase.

In addition, our dataset includes a uniquely valuable variable: the valuation of each firm as of 2024. This valuation reflects external investors’ assessments of firm assets and operational performance and typically serves as a key indicator for determining the next round of funding. It is also widely regarded as a reliable proxy for a firm's overall financial and strategic standing. Accordingly, in the subsequent empirical analysis, we incorporate firm valuation as a testing variable to help isolate the effect of innovation and to further separate it from the signaling function that patents may otherwise serve.

\begin{figure}[htbp]
    \centering
    \includegraphics[width=0.3\textwidth]{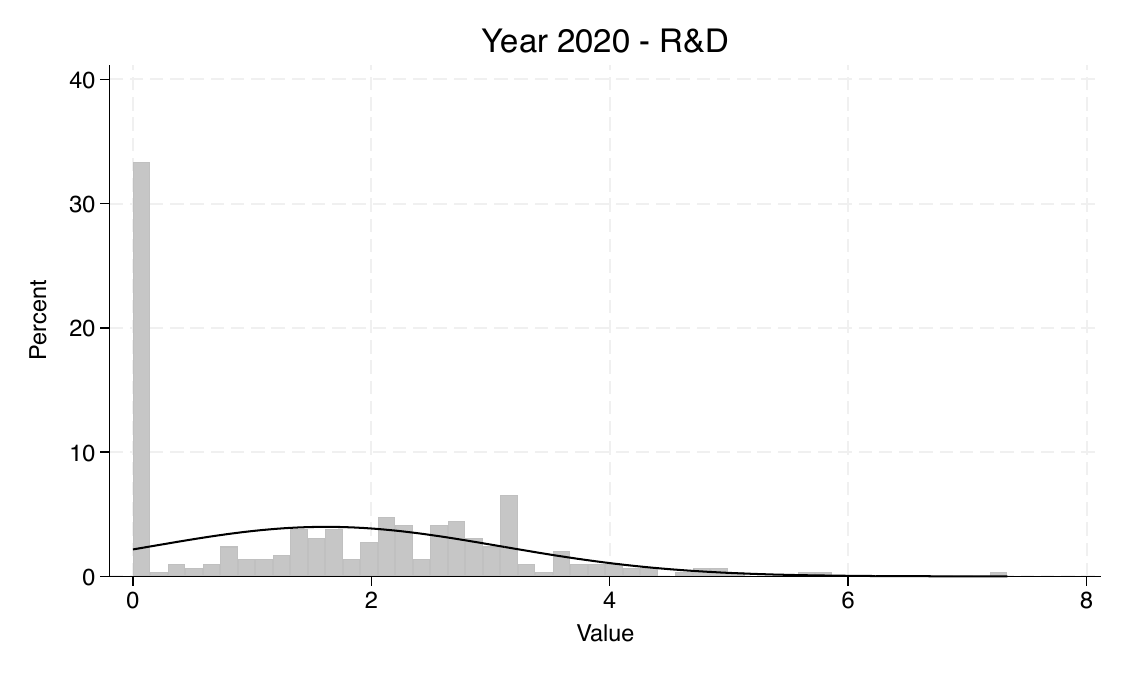}
    \includegraphics[width=0.3\textwidth]{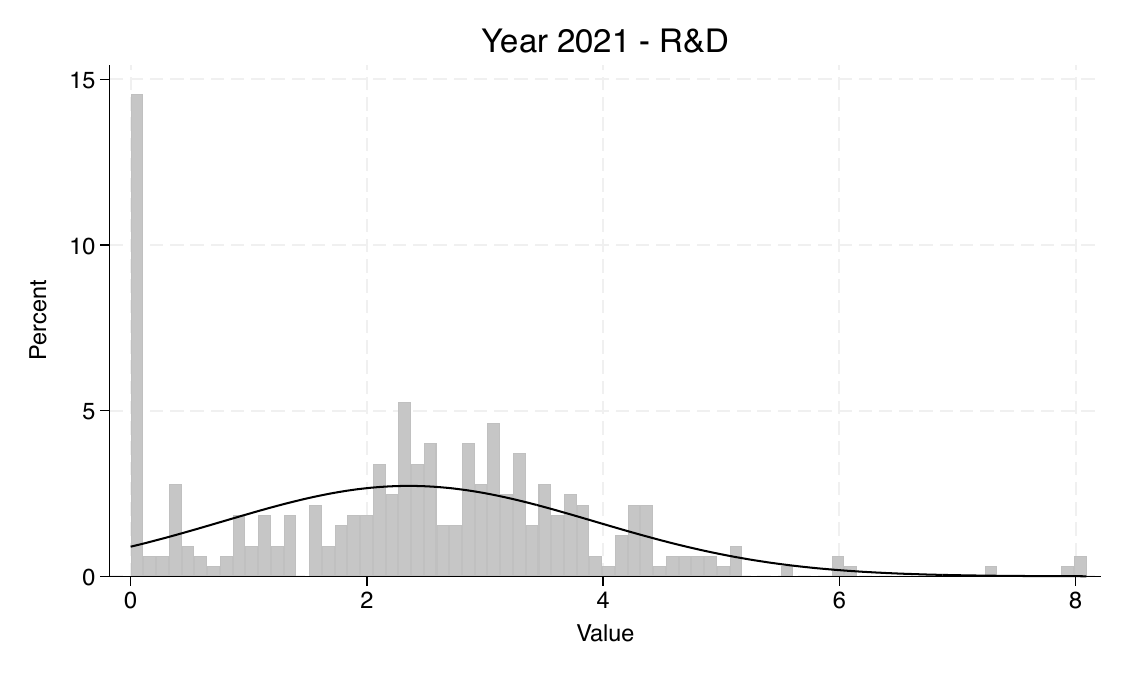}
    \includegraphics[width=0.3\textwidth]{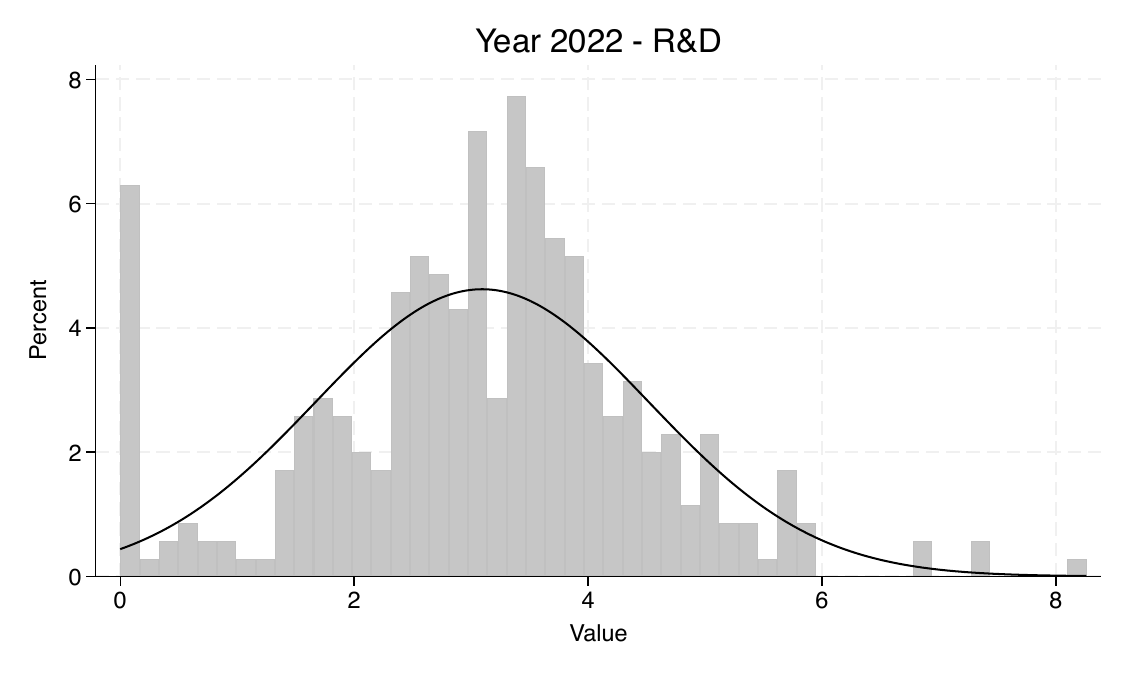}
    
    \vspace{0.2cm}
    \includegraphics[width=0.3\textwidth]{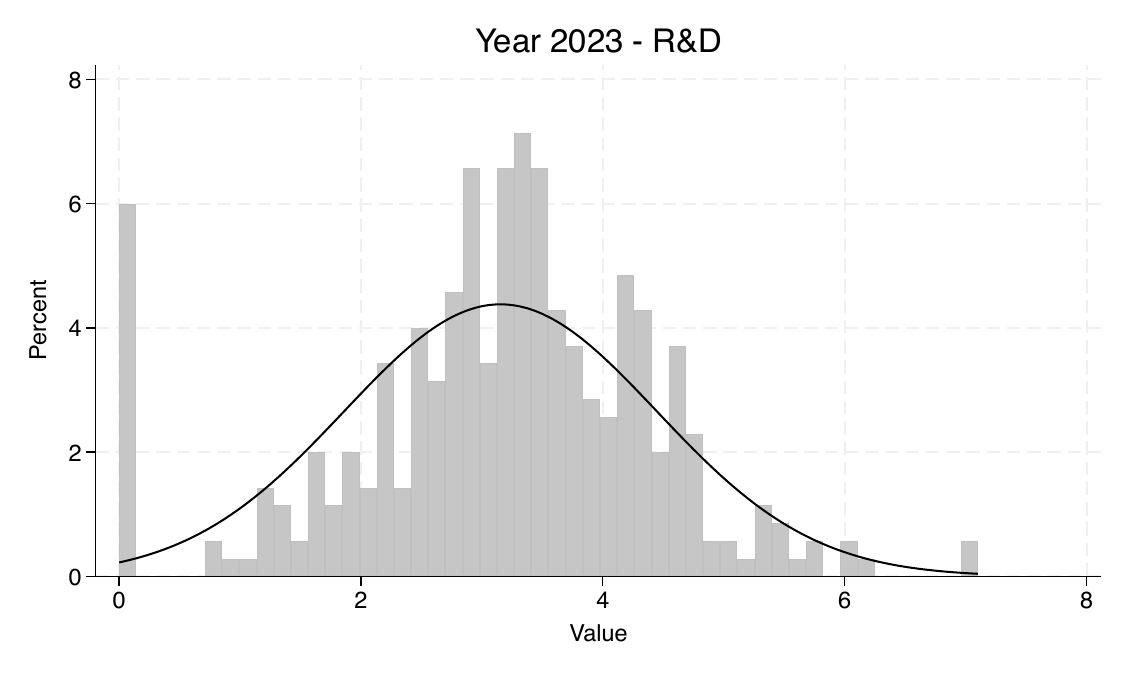}
    \includegraphics[width=0.3\textwidth]{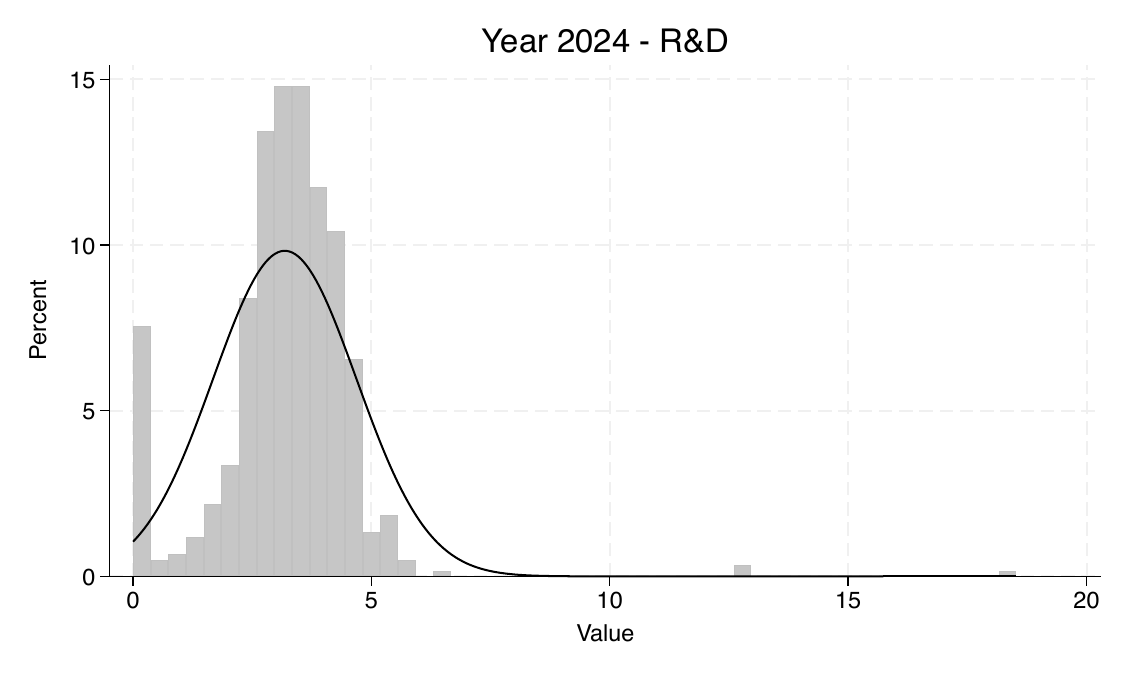}

    \caption{Distribution of R\&D per Employee (2020–2024)}
    \label{fig:rd_years}
\end{figure}

\begin{figure}[htbp]
    \centering
    \includegraphics[width=0.75\textwidth]{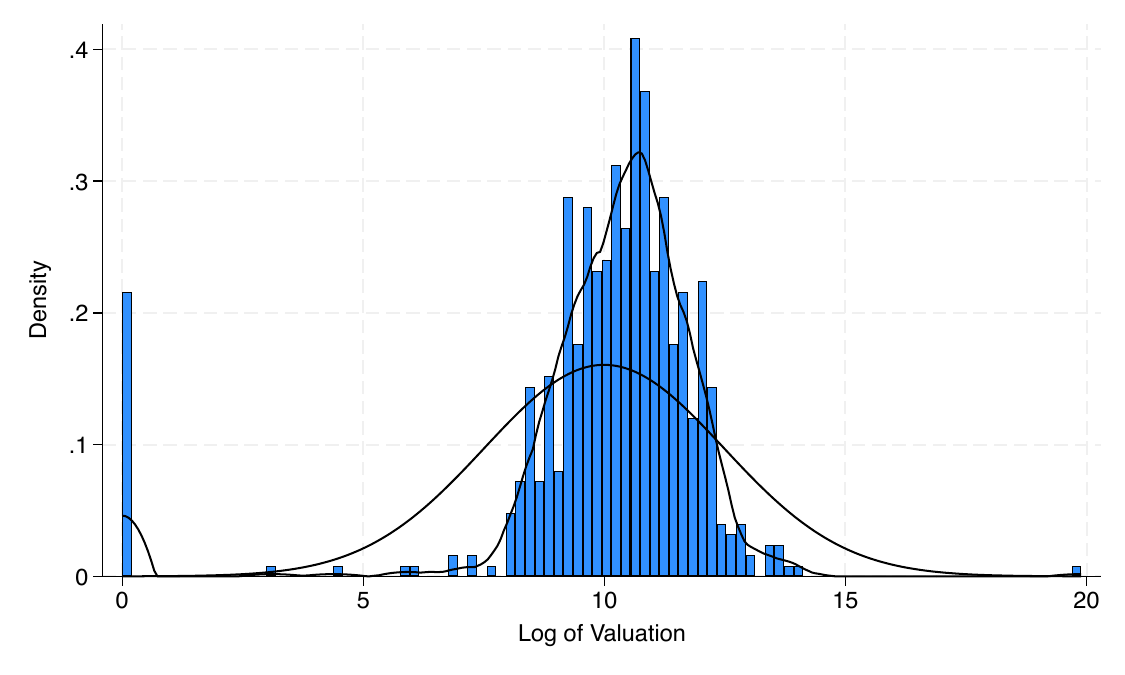}
    \caption{Density Plot of Valuation}
    \label{fig:valuation_density}
\end{figure}

Regarding our core regression variables:

\begin{itemize}
    \item \textbf{Patents}: We use multi-year patent counts, categorized into precise types (e.g., utility model patents, invention patents, and PCT patents). To standardize across firms of varying size, we scale patent counts by the number of employees.
    \item \textbf{Productivity and R\&D}: We use multi-year R\&D investment data and productivity metrics, both of which are log-transformed to reduce skewness, address outliers, and mitigate the risk of over-investment or over-valuation bias (see more in Figure~\ref{fig:rd_years}).
    \item \textbf{Valuations}: Also log-transformed for consistency and comparability (see Figure~\ref{fig:valuation_density}).
\end{itemize}

The productivity index comprises multiple variables that are rigorously examined in our main regressions and robustness checks. These specifications, along with our normalization methods and log transformations, ensure statistical validity and comparability across firms and years.

\section{Theoretical Framework and Empirical Specification}

\subsection{Conceptual Framework}
We follow the standard approach to analyze firm productivity by linking inputs and outputs within a production function framework \citep{CDfunction}. For simplicity, we assume that the functional form of the production function follows a Cobb-Douglas form. We begin with a standard specification relating a firm’s output to its capital and labor inputs:
\begin{equation}
Y_{it} = A_{it} \cdot K_{it}^{\alpha} \cdot L_{it}^{\beta}
\end{equation}
where $Y_{it}$ denotes the productivity of firm $i$ at time $t$. Due to data limitations, direct measures of productivity are typically unavailable in practice; hence, we adopt two proxy variables: pure profit and income. $K_{it}$ represents capital input, $L_{it}$ indicates labor input, and $A_{it}$ captures total factor productivity (TFP), which reflects both a firm’s innovative capacity and other unobserved efficiency components.

Taking logs of both sides gives:
\begin{equation}
\ln Y_{it} = \ln A_{it} + \alpha \ln K_{it} + \beta \ln L_{it}
\end{equation}

To isolate the productivity component $A_{it}$, we assume it is a function of innovation-related variables such as patent output, R\&D investment, and other firm-level controls:
\begin{equation}
\ln A_{it} = \theta_1 \cdot \text{Patent}_{it} + \theta_2 \cdot \text{R\&D}_{it} +  \mu_i + \lambda_t + \varepsilon_{it}
\end{equation}

In our empirical design, patents are incorporated not merely as innovation indicators, but as \textbf{realized outcomes of the innovation process}—a central explanatory variable that reflects a firm’s success in transforming inputs into protectable technological advances. They are not abstract proxies but actual, discrete results that emerge from internal innovation efforts. Given their observable, countable nature and strategic relevance in technology-driven industries, patents offer a concrete manifestation of innovation output, especially for early-stage firms that lack long commercialization pipelines.

To isolate the effect of such innovation outputs, we treat R\&D investment as a \textbf{core control variable} that captures the magnitude of innovation input. A substantial body of prior research affirms the foundational link between R\&D and patenting. For instance, \citet{https://doi.org/10.1111/j.1540-5885.2010.00747.x} documents a robust and persistent correlation between R\&D expenditures and patent production across firm types and industries. This aligns with broader literature suggesting that internal research capabilities are a critical driver of total factor productivity ($A_{it}$ in our formulation). Conceptually, R\&D represents the investment in developing novel technologies, while patents capture the realized inventive output. Thus, controlling for R\&D enables us to separate \textbf{innovative effort} from \textbf{innovative success}, offering a more nuanced understanding of productivity dynamics.

However, a core econometric challenge arises from \textbf{omitted variable bias and reverse causality}. As emphasized by \citet{ARORA20081153}, R\&D and patenting are not exogenously assigned; rather, they are jointly determined by latent factors such as founder experience, strategic orientation, technological opportunity, or access to external resources. We believe that most of these confounding variables are absorbed by the fixed effects included in the model. Notably, a firm’s investment base may influence both investment intensity and innovation outcomes (i.e., patents and R\&D), thereby confounding their observed relationship with productivity. This issue is particularly salient for early-stage ventures, where data on investment rounds and allocation are often incomplete or inconsistently reported.

However, we argue that if general investment is to affect patenting outcomes, it must operate through the channel of R\&D, since the patents in our dataset are not acquired externally but generated internally through firms' own innovation processes. Therefore, R\&D can be viewed as a \textbf{reduced-form proxy} that partially captures the effects of underlying investment flows. Moreover, as noted by the \citet{ft_bigtech_invest2024}, they underscored the centrality of R\&D relative to capital investment among high-tech firms. In early-stage, innovation-driven firms, R\&D---rather than general capital investment---is the primary mechanism through which financial resources influence innovation outcomes. Controlling for R\&D investment, therefore, helps mitigate potential bias arising from capital heterogeneity when estimating the impact of patents on productivity. This strategy also aligns with our broader effort to isolate and control for the signaling function of patents.

Another important clarification concerns the capital data we use. Due to data limitations, it is challenging to obtain precise measures of capital stock for each year. Fortunately, we were able to access annual asset data, along with yearly cost and net profit figures. To approximate annual capital, we manually calculated depreciation-adjusted capital by incorporating industry-specific depreciation rates, as provided by Wind, across different sub-industries (see Table~\ref{tab:industry_code_eng} for details). While this approach is not perfect, we believe it provides a reasonable and consistent estimation of capital under the constraints of the available data.

Substituting Equation (3) into Equation (2), we obtain the following estimable equation:
\begin{equation}
\ln Y_{it} = \theta_1 \cdot \text{Patent}_{it} + \theta_2 \cdot \text{R\&D}_{it} + \alpha \ln K_{it} + \beta \ln L_{it} + \mu_i + \lambda_t + \varepsilon_{it}
\end{equation}
where $\mu_i$ represents firm-specific fixed effects and $\lambda_t$ captures year-specific shocks.

\subsection{Standardized Form: Per-Employee Productivity}

To better reflect productivity across firms of varying size, we further standardize the model by expressing both outputs and inputs on a per-employee basis. This yields:
\begin{equation}
\ln\left(\frac{Y_{it}}{L_{it}}\right) = \theta_1 \cdot \text{Patent}_{it} ^{\prime} + \theta_2 \cdot \text{R\&D}_{it} ^{\prime} +  \alpha  \ln\left(\frac{K_{it}}{L_{it}}\right) + \mu_i^{\prime} + \lambda_t^{\prime} + \varepsilon_{it}
\end{equation}

This form allows for interpretation of innovation impact on labor productivity, where:
\begin{itemize}
    \item $\text{Patent}_{it} ^{\prime}$ captures standardized patent output, controlling for firm scale.
    \item $\text{R\&D}_{it}^{\prime}$ measures R\&D effort per employee, providing a clean input proxy for innovation. To ensure robustness, we will test alternative specifications such as $\text{R\&D}/\text{Profit}$, $\text{R\&D}/\text{Costs}$, and $\text{R\&D}/\text{Revenue}$, following measurement suggestions in \citet{murphy1996measuring}.
    \item $\ln\left(\frac{K_{it}}{L_{it}}\right)$ represents capital intensity per employee.
\end{itemize}

\bigskip

\noindent \textbf{Proposition 1} \textit{Final Theoretical Model.} Based on the derivations above, we simplify our model as:
\begin{equation}
\ln\left(\frac{Y_{it}}{L_{it}}\right) = \beta+ \theta_1 \cdot \text{Patent}_{it}^{\prime} + \theta_2 \cdot \text{R\&D}_{it}^{\prime} + \theta_3 \cdot  \ln\left(\frac{K_{it}}{L_{it}}\right) + \mu_i^{\prime} + \lambda_t^{\prime} + \varepsilon_{it}
\end{equation}

This formulation yields several testable hypotheses: \newline
\textbf{Hypothesis 1}
\label{hypo}

Our key parameter of interest, $\theta_1$, is expected to be greater than zero. This stems from our central argument that, after appropriately controlling for alternative channels such as R\&D, patents serve as a robust proxy for firms' true innovative capacity. As noted by \citet{matricano2020effect}, patents positively affect firm-level innovation outcomes, particularly in high-tech sectors, which closely aligns with the context of our study. Concerns raised in the literature—such as the role of “nominal patents”—are unlikely to be relevant in our setting, which focuses on emerging and innovation-driven enterprises.

We further argue that the commonly cited signaling function of patents, as discussed in \citet{USsoftwarestartups}, can be ruled out for two reasons. First, as we clarified in the data section, all firm observations are from periods after their most recent round of venture capital investment. This eliminates the possibility that patents are primarily used to attract capital in subsequent rounds. These firms are no longer in the stage of seeking funds using patent signaling; rather, they are now leveraging patents to improve substantive firm outcomes. Second, we control for R\&D and other investment-related factors. This reinforces our claim that, in this context, patents primarily capture innovation rather than acting as a signal, which is consistent with our review of literature on high-tech start-ups in China—where innovation value dominates symbolic signaling.

These theoretical expectations guide our empirical framework and inform the structure of our identification strategy, which is designed to account for potential heterogeneity, endogeneity, and omitted variable bias.

\subsection{Empirical Model}

Building on the above, our empirical model is formalized as follows:
\begin{equation}
\label{equation 8}
\ln\left(Y_{it}\right) = \alpha + \beta_1 \cdot \text{Patent}_{it} + \beta_2 \cdot \ln \left(\text{R\&D}_{it}\right) + \beta_3 \cdot \ln \left(\text{Capital}_{it}\right) + \gamma^\top X_{it} + \mu_i + \lambda_t + \varepsilon_{it}
\end{equation}

All variables are standardized to eliminate firm size effects. The dependent variable $Y_{it}$ denotes the performance of firm $i$ in year $t$. In our main specification, this is proxied by income and pure profit. $\text{Patent}_{it}$ represents standardized patent output for firm $i$ in period $t$. 

The vector $X_{it}$ includes several time-varying firm-level indicators that may affect both innovation inputs and outcomes. These include whether the firm is recognized as a national high-tech enterprise, an “innovative SME”, or a specialized, refined, distinctive, and innovative firm at the municipal, provincial, or national level. Such designations may confer differential access to subsidies, tax relief, and government resources \citep{gaoxinservjice}, thereby potentially affecting both R\&D and firm performance.

We include firm fixed effects ($\mu_i$) to control for unobserved, time-invariant heterogeneity such as founder ability, early firm culture, and core technical capacity. Year fixed effects ($\lambda_t$) absorb macroeconomic fluctuations, regulatory shocks, and other time-specific factors common across firms. Additionally, in extended specifications, we incorporate industry fixed effects to control for sectoral productivity norms and R\&D intensities.

This fixed-effects strategy addresses several key methodological challenges. Existing studies of early-stage firms often rely on between-firm comparisons using coarse observables, yet we argue that unobservable founder characteristics and internal firm dynamics exert a first-order influence on innovation behavior and performance outcomes. These latent characteristics also correlate with our key independent variables—such as patenting activity and R\&D intensity—introducing self-selection bias. By focusing on within-firm changes over time, our specification mitigates this concern and yields more credible estimates. Specifically, $\mu_i$ captures stable firm-level traits (e.g., managerial ability, initial funding environment), while $\lambda_t$ accounts for time-specific shocks that are homogeneous across firms.

This framework enables us to isolate the marginal contribution of patents to firm performance—conditional on R\&D input and observed heterogeneity—and to test the extent to which patenting functions as a mechanism of innovation rather than a signal for capital access.

\section{Main Results}
\subsection{Impact of Patents on Income of the Companies}

To examine the impact of innovation outcomes (i.e., patents) on companies’ income, we employ the baseline regression model specified in Equation~\ref{equation 8}. Table~\ref{table8: main results 1} reports the estimated effects of patents on company-level productivity, measured by reported annual income. Both the patent and income variables are scaled by the number of employees to mitigate the confounding effects of firm size. For the income variable, we apply the logarithmic transformation to preserve the original distribution structure while addressing the significant heterogeneity in scale across high-tech firms, as documented in the data section. This log-linear specification also facilitates meaningful economic interpretation of the coefficients.

Column (1) in Table~\ref{table8: main results 1} shows that the coefficient on \textit{Patent} is positive but statistically insignificant. This result is initially counterintuitive, as it contradicts our theoretical expectation that greater innovation output should be associated with higher firm income. The coefficient remains statistically insignificant even after incorporating additional controls, such as whether the firm holds specific policy-recognized certifications, which may entitle them to preferential financing or regulatory treatment (columns 2 and 3). In the fully controlled model in column (3), the coefficient on \textit{Patent} is 0.0046, suggesting that a one-unit increase in patents corresponds to an approximate 0.46\% increase in income—yet this estimate is not statistically significant.

At first glance, this result may appear at odds with our earlier argument that patents, after the signaling function has been carefully controlled for, serve as a valid proxy for innovation. However, the null finding in fact offers empirical support for a more nuanced interpretation: it is precisely because we restrict our sample to the post-VC operational period—thereby minimizing signaling bias—that the remaining effect of patents on income appears weak. This implies that, in many early-stage firms, the income-enhancing role of patents may be largely attributable to their signaling function rather than intrinsic innovation value. In other words, once signaling is accounted for, the standalone impact of patent output on firm performance is substantially diminished.

This insight is further developed in our subsequent sections, where we explicitly test the extent to which patents are shaped by strategic, rather than purely technological, considerations. We examine how firm valuation and policy incentives influence patenting behavior, and whether these institutional factors are more predictive of income than patent counts per se.

It is also worth noting that both \textit{R\&D Input} and \textit{Capital} enter the regression significantly. After logarithmic transformation, these inputs display strong positive effects on income at the 1\% significance level. Specifically, a 1\% increase in R\&D input is associated with a 29\% increase in income, while a 1\% increase in capital corresponds to a 63\% increase in income. These results affirm that genuine investment efforts, rather than patent counts alone, are more directly linked to firm performance—an observation that reinforces our argument regarding the complex, and potentially overstated, role of patents in driving productivity.

{\renewcommand{\arraystretch}{1.4} 
\def\sym#1{\ifmmode^{#1}\else\(^{#1}\)\fi}
\begin{table}[htbp]\centering
\caption{How Patents Influence Incomes of Companies}
\hspace{-0.3cm}
\label{table8: main results 1}
\begin{tabular}{l@{\hskip 1.5em}c@{\hskip 1.5em}c@{\hskip 1.5em}c}
\hline\hline
 & (1) & (2) & (3) \\
 & Income & Income & Income \\
\hline
Patent & 0.0049 & 0.0046 & 0.0046 \\
 & (0.0050) & (0.0046) & (0.0047) \\
R\&D & 0.3011\sym{***} & 0.2908\sym{***} & 0.2903\sym{***} \\
 & (0.0924) & (0.0978) & (0.0977) \\
Capital & 0.6764\sym{***} & 0.6305\sym{***} & 0.6300\sym{***} \\
 & (0.1010) & (0.1051) & (0.1052) \\
``Innovative SME" & 0.6275\sym{***} & 0.4466\sym{**} & 0.4454\sym{**} \\
 & (0.1743) & (0.1846) & (0.1846) \\
``Giant Companies" & 0.5220\sym{***} & 0.4094\sym{***} & 0.3587\sym{***} \\
 & (0.1294) & (0.1254) & (0.1362) \\
Municipal Innovative Firm &  & 0.1694 & 0.1664 \\
 &  & (0.1884) & (0.1885) \\
Provincial Innovative Firm &  & 0.6375\sym{***} & 0.6325\sym{***} \\
 &  & (0.1445) & (0.1449) \\
National Innovative Firm &  &  & 0.1732 \\
 &  &  & (0.2270) \\
Constant & -1.2254\sym{***} & -1.0680\sym{***} & -1.0644\sym{***} \\
 & (0.3761) & (0.3896) & (0.3902) \\
Observations & 1,011 & 1,011 & 1,011 \\
R-squared & 0.417 & 0.440 & 0.440 \\
Number of firm\_id & 421 & 421 & 421 \\
\hline
\multicolumn{4}{l}{\footnotesize \textit{Robust standard errors in parentheses}} \\
\multicolumn{4}{l}{\footnotesize \textit{*** p$<$0.01, ** p$<$0.05, * p$<$0.1}} \\
\multicolumn{4}{l}{\footnotesize \textit{All the variables have been normalized as pre-employee.}} \\
\hline\hline
\end{tabular}
\end{table}
}

\subsection{Impact of Patents on Pure Profit of the Companies}

As shown in Table~\ref{table9: main results 2}, we are again surprised to find that patents exert an insignificant effect on firms’ performance, this time measured by pure profit. This finding mirrors the earlier result in Table~\ref{table8: main results 1}, where patents also failed to significantly affect income, reinforcing our broader conclusion that patent counts alone may not be a robust driver of firm-level financial outcomes.

Notably, \textit{R\&D Input} remains significant at the 5\% level, while \textit{Capital}—previously significant in the income regression—loses statistical significance in this setting. While further explanation is provided in the following sections, this consistent insignificance of the patent variable across two key performance metrics suggests that patents may not play as substantial a role in firm success as commonly assumed.

An additional point of concern may arise regarding the different roles of capital across these two models. Specifically, capital is significant in explaining income but not pure profit. We argue that this discrepancy is not indicative of data inconsistency but rather reflects the underlying economic structure of production. In accordance with the Cobb–Douglas production function, capital serves as one of the fundamental inputs contributing to gross output. Firms with larger capital stock typically generate higher income, simply due to a broader operational base. However, these firms also tend to incur higher depreciation and fixed costs, which directly affect their net profit. Therefore, it is not surprising that capital is significantly associated with income but not with pure profit, as the latter captures the residual return after subtracting operating and capital-related costs. This interpretation supports our model design and aligns with theoretical expectations regarding capital's dual role in scale and cost.

{\renewcommand{\arraystretch}{1.4} 
\def\sym#1{\ifmmode^{#1}\else\(^{#1}\)\fi}
\begin{table}[htbp]\centering
\vspace{-2.5cm}
\caption{How Patents Influence Pure Profit of Companies}
\hspace*{-0.5cm}
\label{table9: main results 2}
\begin{tabular}{l@{\hskip 1.5em}c@{\hskip 1.5em}c@{\hskip 1.5em}c}
\hline\hline
 & (1) & (2) & (3) \\
 & Pure Profit & Pure Profit & Pure Profit \\
\hline
Patent & 0.3520 & 0.4042 & 0.3766 \\
 & (0.2777) & (0.2988) & (0.3080) \\
R\&D & 0.6764\sym{**} & 0.6639\sym{**} & 0.5851\sym{**} \\
 & (0.2868) & (0.2791) & (0.2919) \\
Capital & 0.1420 & 0.1448 & 0.1839 \\
 & (0.1682) & (0.1571) & (0.1637) \\
``Innovative SME" & -0.8138 & -1.1517 & -1.0933 \\
 & (0.9453) & (1.0268) & (1.0415) \\
``Giant Companies" & -0.1276 & -0.1280 & -0.4944 \\
 & (0.6986) & (0.7077) & (0.4779) \\
Municipal Innovative Firm &  & 1.8277 & 1.8295 \\
 &  & (1.2645) & (1.2472) \\
Provincial Innovative Firm &  & -0.4707 & -0.4798 \\
 &  & (0.8889) & (0.8895) \\
National Innovative Firm &  &  & 1.2529\sym{***} \\
 &  &  & (0.3834) \\
Constant & -0.7319 & -0.7768 & -0.7577 \\
 & (0.6980) & (0.7060) & (0.7054) \\
Observations & 141 & 141 & 141 \\
R-squared & 0.272 & 0.321 & 0.337 \\
Number of firm\_id & 87 & 87 & 87 \\
\hline
\multicolumn{4}{l}{\footnotesize \textit{Robust standard errors in parentheses}} \\
\multicolumn{4}{l}{\footnotesize \textit{*** p$<$0.01, ** p$<$0.05, * p$<$0.1}} \\
\multicolumn{4}{l}{\footnotesize \textit{All the variables have been normalized as pre-employee.}} \\
\hline\hline
\end{tabular}
\end{table}
}

\section{Mechanism and Explanations}
\subsection{Why Patents Don't Take Effect as We Expect?}

Patents, as typical outputs of innovation activities, are widely documented in the literature to enhance firm productivity (see Section~\ref{subsec:mature} for detailed discussion). Given the concerns that such effects may not hold in the context of early-stage firms or may be confounded by other factors, we have already proposed several contextual explanations in Sections~\ref{sub: our context} and~\ref{hypo}. In this section, rather than reiterating those arguments, we aim to revisit and reinforce them using empirical testing.

Two main explanations, frequently discussed in the literature, offer potential insight into our null results:

\begin{enumerate}
    \item \textbf{Patents function primarily as signals rather than as substantive innovation outputs}, allowing start-ups to attract external funding or venture capital (VC).
    \item \textbf{Some patents may not represent genuine innovation}, but instead serve as nominal tools to obtain market advantages or government recognition, thus lacking true technological contribution.
\end{enumerate}

\vspace{0.5em}
\textbf{Revisiting the Signaling Hypothesis.} As discussed in Section~\ref{data part}, we attempt to control for signaling effects by limiting our sample to the period between the first VC investment and the most recent external financing, thus ensuring that observed patenting activity follows the initial investment. This design reduces the likelihood that patents are used as ex ante signals to attract initial VC. However, we acknowledge that some firms may still seek subsequent funding rounds even within this short time window. Hence, we cannot completely rule out the possibility that patenting behavior continues to reflect signaling motives aimed at future investors.

To empirically examine this residual signaling effect, we conduct two complementary tests. The first test explores the impact of patents on firms’ valuation in 2024, which serves as a proxy for external investor perception and next-round funding expectations \citep{damodaran2012investment}. Table~\ref{tab: signal1} presents the regression results. Across all specifications, we find that \textit{patents per employee} significantly and positively influence firm valuation. For instance, in the most comprehensive model (column 3), each additional patent per employee corresponds to a 0.108 increase (in 10,000 Yuan) in firm valuation. These results remain stable when controlling for R\&D input and capital, suggesting that patents still convey meaningful signals to external stakeholders, even after accounting for other innovation-related variables.

This implies that companies may choose to issue patents not primarily for their immediate productivity effects, but rather for their anticipated signaling value in future financing rounds. In other words, patenting may serve as a strategic action with lagged benefits via increased valuation, rather than delivering short-term operational gains.

The second test adopts a robustness approach by analyzing the relationship between patents and the \textit{total amount} and \textit{number of subsequent investment rounds} following the first VC investment. Results in Table~\ref{tab: signal2} reveal a similar pattern: patents remain a significant predictor of follow-up financing, reinforcing the notion that their role as signals persists even after initial investment. While our data does not allow precise timing of each funding event, the aggregate influence is sufficiently clear to support our interpretation.

Together, these two empirical exercises provide strong support for the signaling explanation. Even within a period designed to reduce pre-investment signaling bias, patents continue to influence variables—valuation and subsequent funding—that are central to investment decisions. This may partially explain why patents fail to significantly impact income or profit in our baseline regressions: their functional role may lie outside the productivity channel.

\begin{table}[htbp]
\centering
\caption{Do Patents Signal to Investors? (Testing 1)}
\begin{threeparttable}
  {
\def\sym#1{\ifmmode^{#1}\else\(^{#1}\)\fi}
\begin{tabular}{l*{3}{c}}
\hline\hline
                    &\multicolumn{1}{c}{(1)}&\multicolumn{1}{c}{(2)}&\multicolumn{1}{c}{(3)}\\
                    &\multicolumn{1}{c}{valuation}&\multicolumn{1}{c}{valuation}&\multicolumn{1}{c}{valuation}\\
[0.5em]
\hline
Total Patents    &       0.160\sym{***}&       0.119\sym{***}&       0.108\sym{***}\\
                    &    (0.0282)         &    (0.0297)         &    (0.0339)         \\
[1em]
Capital &  0.00000262\sym{**} &  0.00000954\sym{***}&  0.00000964\sym{***}\\
                    &(0.00000127)         &(0.00000188)         &(0.00000181)         \\
[1em]
Income in 2024           &-0.000000518         &  0.00000615\sym{***}&  0.00000496\sym{***}\\
                    &(0.00000277)         &(0.00000153)         &(0.00000175)         \\
[1em]
R\&D input in 2024      &       0.209\sym{***}&       0.178\sym{***}&       0.153\sym{***}\\
                    &    (0.0441)         &    (0.0401)         &    (0.0376)         \\
[1em]
Employee Numbers   &     &      -0.00437\sym{***}&      -0.00472\sym{***}\\
                    &       &    (0.000600)         &    (0.000668)         \\
[1em]
Which Rounds &  & \checkmark & \checkmark \\
[1em]
Industry FE &  & \checkmark & \checkmark \\
[1em]
Year FE     &  &  & \checkmark \\
[1em]
Stage FE    &  &  & \checkmark \\
[1em]
SZ Indicator &  &  & \checkmark \\
[0.5em]
\hline
Observations        &         583         &         583         &         478         \\
[0.5em]
R-squared           &       0.153         &       0.267         &       0.294         \\
\hline\hline
\end{tabular}
}

  \begin{tablenotes}[flushleft]
    \footnotesize
    \item[] \textit{Notes:} 
    \item[] \hspace{0.2cm} Log transformations are applied to capital, income, and R\&D input for consistency with previous specifications. Patents are kept in levels due to their relatively small magnitude.
    \item[] \hspace{0.2cm} All variables are normalized by employee count. We also include employee size directly in the regression to capture team scale effects. Interestingly, employee size appears negatively associated with valuation, possibly reflecting inefficiencies in large teams at early stages.
    \item[] \hspace{0.2cm} ***$p<0.01$, **$p<0.05$, *$p<0.1$.
  \end{tablenotes}
\end{threeparttable}
\label{tab: signal1}
\end{table}

\begin{table}[htbp]
\centering
\caption{Do Patents Signal to Investors? (Testing 2)}
\begin{threeparttable}
  {
\def\sym#1{\ifmmode^{#1}\else\(^{#1}\)\fi}
\begin{tabular}{l*{3}{c}}
\hline\hline
                    &\multicolumn{1}{c}{(1)}&\multicolumn{1}{c}{(2)}&\multicolumn{1}{c}{(3)}\\
                    &\multicolumn{1}{c}{Sunsequent Inv.}&\multicolumn{1}{c}{Sunsequent Inv.}&\multicolumn{1}{c}{Sunsequent Inv.}\\
[0.5em]
\hline
Total Patents   &       0.228\sym{***}&       0.140\sym{***}&       0.124\sym{***}\\
                    &    (0.0431)         &    (0.0361)         &    (0.0373)         \\
[1em]
Capital & -0.00000969\sym{***}&  0.00000476\sym{*}  &  0.00000480\sym{*}  \\
                    &(0.00000276)         &(0.00000288)         &(0.00000247)         \\
[1em]
Total Income    &   0.0000462\sym{***}&   0.0000457\sym{***}&    0.000526\sym{***}\\
                    &(0.00000490)         &(0.00000487)         &  (0.000126)         \\
[1em]
Total R\&D Input       &       0.124\sym{**} &      0.0958\sym{**} &      0.0937\sym{*}  \\
                    &    (0.0528)         &    (0.0441)         &    (0.0513)         \\
[1em]
Employee Numbers   &                     &    -0.00733\sym{***}&    -0.00706\sym{***}\\
                    &                     &  (0.000682)         &  (0.000612)         \\
[1em]
Which Rounds &  & \checkmark & \checkmark \\
[1em]
Industry FE &  & \checkmark & \checkmark \\
[1em]
Year FE     &  &  & \checkmark \\
[1em]
Stage FE    &  &  & \checkmark \\
[1em]
SZ Indicator &  &  & \checkmark \\
[0.5em]
\hline
Observations        &         325         &         325         &         268         \\
[0.5em]
R-squared           &       0.351         &       0.683         &       0.709         \\
\hline\hline
\end{tabular}
}

  \begin{tablenotes}[flushleft]
    \footnotesize
    \item[] \textit{Notes:} 
    \item[] \hspace{0.2cm} The model mirrors Table~\ref{tab: signal1}, but replaces valuation with cumulative post-first-round investment data. Due to data constraints, investment timing cannot be precisely identified.
    \item[] \hspace{0.2cm} All variables are log-transformed where appropriate and normalized by employee count.
    \item[] \hspace{0.2cm} ***$p<0.01$, **$p<0.05$, *$p<0.1$.
  \end{tablenotes}
\end{threeparttable}
\label{tab: signal2}
\end{table}

\vspace{0.5em}
\textbf{Reinterpreting the ``Nominal" Patent Hypothesis.} 

As for the second explanation—that some patents may be nominal rather than substantive—we maintain our original skepticism regarding its prevalence in high-tech start-ups. Following \citet{ahuja2001entrepreneurship}, nominal patents are characterized by strategic but non-innovative intent, often driven by traps such as familiarity, maturity, or proximity. We argue that such traps are less relevant in dynamic start-up contexts, where the pursuit of novel technology is often essential for survival.

Nonetheless, we propose a variant of the nominality hypothesis rooted in the institutional environment of Chinese high-tech start-ups. As emphasized in \citet{gaoxinservjice}, various government support programs—such as the ``Innovative SME'' and ``National High-Tech Enterprise'' recognitions—use patent counts as eligibility criteria. These designations offer tangible policy benefits, including tax exemptions and preferential financing. Given this incentive structure, firms may strategically increase patent output to qualify for such recognitions, regardless of the underlying innovation value.

Due to limited data access, we are unable to directly test this hypothesis. However, our main regression results (Tables~\ref{table8: main results 1} and~\ref{table9: main results 2}) show that firms holding these recognitions exhibit significantly higher income and profit. This suggests that policy-related incentives may partially mediate the effect of patents on firm performance—absorbing what would otherwise appear as patent-driven effects.

In this context, patents may serve a “nominal policy” function: they act as prerequisites for recognition, while the actual performance benefits are realized via state support mechanisms. This offers a complementary explanation for why patents fail to exhibit significant direct effects on productivity in our baseline models.

\vspace{0.5em}
\textbf{Summary.} In summary, our mechanism analysis suggests that patents in early-stage high-tech firms are often instrumentalized—either as signals to investors or as tools to secure government recognition—rather than serving as intrinsic innovation drivers. This recontextualizes the null results found earlier and underscores the importance of understanding the multifaceted roles patents play in entrepreneurial ecosystems.

\subsection{How Can R\&D Take Effect on Productivity?}

Unlike patents, which represent outcomes of innovation, R\&D captures the \textit{process} and investment in innovation activities. It aligns with common sense that innovation input can drive productivity, but to better understand this effect, we aim to answer two guiding questions: 

\begin{enumerate}
    \item Through which channels does R\&D exert influence on firm-level productivity?
    \item What are the key forces driving R\&D investment in high-tech start-ups?
    \item Does R\&D investment effectively translate into patent output in early-stage high-tech firms?
\end{enumerate}

\vspace{0.5em}
\textbf{(1.1) Mechanism 1: Organizational Environment Enhancement.}

Although we lack direct observational data on firms’ internal environments, there is extensive theoretical support for the argument that R\&D improves organizational capabilities. According to \citet{tang2019r}, R\&D intensity is often associated with an innovation-oriented managerial style and organizational flexibility, both of which contribute to long-term firm performance. Similarly, \citet{entwistle1999exploring} argue that sustained R\&D efforts foster a supportive workplace environment, particularly in high-tech firms. Thus, R\&D may indirectly enhance productivity by improving managerial quality and workplace culture.

\vspace{0.5em}
\textbf{(1.2) Mechanism 2: Attracting High-Quality Human Capital.}

R\&D investment also signals a company’s commitment to technological advancement, potentially attracting top-tier researchers and high-skilled employees. To test this channel empirically, we construct a proxy for employee quality using several indicators: number of employees with PhDs, number of full-time academic staff, and number of researchers with high citation records (according to the records on Web of Science). We aggregate these into an “employee productivity index” by dividing the total number of high-skill employees by total employment.

Table~\ref{tab:rdtesting} reports the regression results. Contrary to our expectation, R\&D input does not exhibit a significant relationship with employee quality. One possible reason is that high-end talents choose their employers based on personalized negotiations and other non-financial factors rather than R\&D intensity per se. Additionally, some R\&D expenditures may be allocated to equipment or outsourced research services, and thus may not directly reflect hiring quality. These findings suggest that the human capital attraction channel may not be a dominant mechanism through which R\&D drives productivity in this context.

However, even if our underlying logic appears sound, an important limitation must be acknowledged. Due to data constraints, we cannot determine the precise timing of R\&D expenditure relative to the recruitment of high-skill employees. It is possible---and even likely---that these employees joined the firms \textit{before} the recorded R\&D investments were made. If so, our regression would fail to capture any causal relationship between R\&D spending and the attraction of human capital, rendering the test inconclusive. This timing mismatch substantially limits the interpretability of our results. Future research should revisit this mechanism with richer datasets that allow for more granular matching of hiring events and R\&D allocation over time.

\begin{table}[htbp]
\centering
\caption{Is R\&D Input Associated with Higher Employee Quality?}
\hspace*{-0.5cm}
\begin{threeparttable}
  {
\def\sym#1{\ifmmode^{#1}\else\(^{#1}\)\fi}
\begin{tabular}{l*{3}{c}}
\hline\hline
                    &\multicolumn{1}{c}{(1)}&\multicolumn{1}{c}{(2)}&\multicolumn{1}{c}{(3)}\\
                    &\multicolumn{1}{c}{Employees' Ratio}&\multicolumn{1}{c}{Employees' Ratio}&\multicolumn{1}{c}{Employees' Ratio}\\
[0.5em]
\hline
R\&D Input       &     0.00880         &     0.00804         &     0.00939         \\
                    &   (0.00772)         &   (0.00750)         &   (0.00862)         \\
[1em]
Capital in 2024         &     0.00616         &     0.00828         &      0.0196\sym{**} \\
                    &   (0.00856)         &   (0.00838)         &   (0.00918)         \\
[1em]
Total Income    & -0.00000353\sym{**} & -0.00000404\sym{**} &   -0.000131\sym{***}\\
                    &(0.00000179)         &(0.00000197)         & (0.0000500)         \\
[1em]
Total Patents' Num.   &     0.00247         &     0.00157         &     0.00395         \\
                    &   (0.00508)         &   (0.00547)         &   (0.00718)         \\
[1em]
Subsequent Inv.     &      0.0301\sym{**} &      0.0340\sym{**} &      0.0405\sym{**} \\
                    &    (0.0140)         &    (0.0149)         &    (0.0159)         \\
[1em]
Which Rounds &  & \checkmark & \checkmark \\
[1em]
Industry FE &  & \checkmark & \checkmark \\
[1em]
Year FE     &  &  & \checkmark \\
[1em]
Stage FE    &  &  & \checkmark \\
[1em]
SZ Indicator &  &  & \checkmark \\
[0.5em]
\hline
Observations        &         324         &         324         &         267         \\
[0.5em]
R-squared           &       0.110         &       0.151         &       0.246         \\
\hline\hline
\end{tabular}
}

  \begin{tablenotes}[flushleft]
    \footnotesize
    \item[] \textit{Notes:} All monetary and R\&D variables are log-transformed and normalized by employee count. The benchmark year is 2024. Although income and investment significantly predict employee structure, R\&D input itself does not. This may reflect the fact that high-quality hiring is more directly associated with wage levels or equity options, which are not necessarily captured within the scope of R\&D accounting.
    \item[] \hspace{0.2cm} ***$p<0.01$, **$p<0.05$, *$p<0.1$.
  \end{tablenotes}
\end{threeparttable}
\label{tab:rdtesting}
\end{table}

\vspace{0.5em}
\textbf{(1.3) Mechanism 3: Policy-Driven Signaling.}

R\&D input is also documented as a requirement for accessing government recognition and policy support. As highlighted by \citet{gaoxinservjice2}, firms often report R\&D expenditures as part of their application for designations such as “Innovative SME” or “High-Tech Enterprise,” which come with tax benefits and financial incentives. Similar to the signaling function of patents discussed earlier, R\&D may serve a signaling role toward policymakers. This mechanism helps explain why firms invest in R\&D even when short-term innovation output (e.g., patents) is limited—because the true productivity gains may be realized via policy channels.
 
\textbf{Summary.}

In summary, although we conceptually identify three potential mechanisms through which R\&D may influence firm-level productivity, empirical testing remains highly constrained. Based on official policy documents, industry reports, and field interviews, we are reasonably confident that the third mechanism---policy-driven signaling---plays a meaningful role in motivating R\&D investment in early-stage firms. However, for the first two channels, the lack of detailed organizational and personnel-level data makes it extremely difficult to construct reliable proxies and perform rigorous causal inference. This data limitation is a common challenge in studying startups and should not be underestimated. Given these constraints, we believe our current strategy represents a reasonable empirical compromise under real-world data conditions. Nevertheless, we call upon future researchers to build more granular datasets that can fully capture firm-level dynamics and provide a more comprehensive understanding of how R\&D activities translate into productivity gains.

\vspace{1em}
\textbf{(2) What Drives R\&D Investment in Early-Stage Firms?}

To address this question, we draw on the investment allocation framework of \citet{joglekar2009marketing}, who suggest that start-ups allocate venture funding between marketing and innovation. We extend this argument by proposing that, in the context of Chinese high-tech firms, investment is disproportionately allocated to R\&D.

We formalize this intuition as follows:
\begin{equation}
    \text{R\&D}_{t} = \alpha \cdot \text{Investment}_{t}, \quad \text{where } \alpha > 0
\end{equation}

Using Wave 2024 cross-sectional data, we test this hypothesis. As shown in Table~\ref{tab:rd_results}, both firm valuation and subsequent investment significantly predict R\&D expenditure. In particular, these relationships remain robust under various controls, though the magnitude of significance decreases slightly in the third specification—likely due to fixed effects absorbing part of the variance.

\begin{table}[htbp]
\centering
\caption{Determinants of R\&D Investment (2024)}
\hspace*{-1.3cm}
\begin{threeparttable}
  {
\def\sym#1{\ifmmode^{#1}\else\(^{#1}\)\fi}
\begin{tabular}{l*{3}{c}}
\hline\hline
                    &\multicolumn{1}{c}{(1)}&\multicolumn{1}{c}{(2)}&\multicolumn{1}{c}{(3)}\\
                    &\multicolumn{1}{c}{R\&D Expenditure}&\multicolumn{1}{c}{R\&D Expenditure}&\multicolumn{1}{c}{R\&D Expenditure}\\
[0.5em]
\hline
Investment Rounds&      0.0814         &      0.0975\sym{**} &       0.111\sym{**} \\
                    &    (0.0498)         &    (0.0486)         &    (0.0527)         \\
[1em]
Valuations&       0.516\sym{***}&       0.490\sym{***}&       0.476\sym{**} \\
                    &     (0.192)         &     (0.186)         &     (0.230)         \\
[1em]
Subsequent Inv. Amounts&       0.347\sym{***}&       0.345\sym{***}&       0.352\sym{**} \\
                    &     (0.117)         &     (0.111)         &     (0.141)         \\
[1em]
Industry FE &  & \checkmark & \checkmark \\
[1em]
Year FE     &  &  & \checkmark \\
[1em]
Stage FE    &  &  & \checkmark \\
[1em]
SZ Indicator &  &  & \checkmark \\
[1em]
\hline
Observations        &         341         &         341         &         284         \\
[0.5em]
R-squared           &       0.487         &       0.509         &       0.555         \\
\hline\hline
\end{tabular}
}

  \begin{tablenotes}[flushleft]
    \footnotesize
    \item[] \textit{Notes:} 
    \item[] All variables are log-transformed and normalized by employee count. Industry, registration year, firm stage, and Shenzhen location fixed effects are included as indicated. 
    \item[] Higher valuation and external capital access appear to positively incentivize R\&D, consistent with the interpretation that R\&D is a primary vehicle through which capital is transformed into innovation capacity.
    \item[] \hspace{0.2cm} ***$p<0.01$, **$p<0.05$, *$p<0.1$.
  \end{tablenotes}
\end{threeparttable}
\label{tab:rd_results}
\end{table}

These findings lead to two core insights. First, firm valuation significantly predicts R\&D input, suggesting that investor expectations shape innovation decisions. Firms anticipating higher future growth tend to allocate more capital toward R\&D. Second, subsequent investment amounts also significantly drive R\&D spending, implying that firms actively reinvest external funds into innovation and innovation is driven by the outside investment.

\vspace{0.5em}
\textbf{(3) R\&D Does Not Always Translate into Patents.}

To further investigate the productivity of R\&D investment, we test whether R\&D intensity translates into increased patent output. As shown in Table~\ref{tab:relation_patent_Capitial_new2}, we find no robust or statistically significant relationship between R\&D and patent counts in Wave 2024. Surprisingly, subsequent investment is negatively associated with patent production. This result complements our earlier arguments that patents may reflect signaling or nominal motives, and that once firms secure sufficient capital, they may have less incentive to continue patenting—especially if patents are not the core vehicle for innovation.

Another plausible explanation aligns with our earlier interpretation of patents as signals rather than substantive innovation outputs. If patents primarily serve a signaling function, then the true innovation activity may still occur through R\&D, which captures the actual investment in knowledge creation. However, firms may strategically choose not to disclose these innovations through patents. Prior literature has documented such behavior: several studies indicate that even when innovations are patentable, firms often refrain from patenting due to perceived costs or the limited marginal benefits associated with disclosure. For instance, \citet{hall2012uk} observe that “some firms avoid the patent system altogether, either because of its cost or because patenting is perceived to yield no additional benefit.” This perspective helps explain the observed disconnect between R\&D intensity and patent output in our data and is consistent with our broader argument regarding the symbolic role of patents in early-stage high-tech firms.

\begin{table}[htbp]
\centering
\caption{R\&D and Capital Show Weak Correlation with Patent Output (2024)}
\hspace*{-1.5cm}
\begin{threeparttable}
  {
\def\sym#1{\ifmmode^{#1}\else\(^{#1}\)\fi}
\begin{tabular}{l*{3}{c}}
\hline\hline
                    &\multicolumn{1}{c}{(1)}&\multicolumn{1}{c}{(2)}&\multicolumn{1}{c}{(3)}\\
                    &\multicolumn{1}{c}{2025 Patents Num.}&\multicolumn{1}{c}{2025 Patents Num.}&\multicolumn{1}{c}{2025 Patents Num.}\\
[0.5em]
\hline
R\&D Expenditure in 2024      &       0.926         &       0.974         &       0.628         \\
                    &     (0.645)         &     (0.653)         &     (0.761)         \\
[1em]
Valuation   &       0.839         &       0.795         &       1.403         \\
                    &     (1.033)         &     (1.037)         &     (1.548)         \\
[1em]
Subsequent Inv.     &      -5.082\sym{***}&      -5.032\sym{***}&      -6.607\sym{***}\\
                    &     (1.580)         &     (1.596)         &     (2.194)         \\
[1em]
Rounds of Inv. &       1.382         &       1.273         &       0.445         \\
                    &     (1.103)         &     (1.146)         &     (1.262)         \\
[1em]
Capital in 2024          &       2.345\sym{***}&       2.552\sym{***}&       3.242\sym{**} \\
                    &     (0.901)         &     (0.947)         &     (1.280)         \\
[1em]
Industry FE &  & \checkmark & \checkmark \\
[1em]
Year FE     &  &  & \checkmark \\
[1em]
Stage FE    &  &  & \checkmark \\
[1em]
SZ Indicator &  &  & \checkmark \\
[0.5em]
\hline
Observations        &         231         &         231         &         190         \\
[0.5em]
R-squared           &       0.152         &       0.199         &       0.309         \\
\hline\hline
\end{tabular}
}

  \begin{tablenotes}[flushleft]
    \footnotesize
    \item[] \textit{Notes:} 2024 is selected as the reference year for aligning capital input and patent output. Valuation and investment data are used as proxies for internal and external capital availability. While R\&D and valuation positively relate to firm growth, they do not necessarily manifest in short-term patent output. This supports the notion that patents may not fully capture innovation productivity in early-stage firms.
    \item[] \hspace{0.2cm} ***$p<0.01$, **$p<0.05$, *$p<0.1$.
  \end{tablenotes}
\end{threeparttable}
\label{tab:relation_patent_Capitial_new2}
\end{table}

\vspace{0.5em}
\textbf{Summary.} 

Overall, we identify three channels through which R\&D may influence productivity: organizational environment enhancement, policy signaling, and capital absorption. However, we find limited evidence for human capital restructuring and for R\&D translating into patent output. Instead, our findings reinforce the idea that R\&D investment is both an endogenous firm decision and a response to investor expectations and policy structures—making it a more consistent and significant driver of productivity than patent counts.

\section{Heterogeneity}

Finally, we explore how the effects of innovation—both in the form of patent output and R\&D input—vary across firms with different characteristics. Specifically, we consider three dimensions of heterogeneity: sub-industry categories, whether the patented technology has a breakthrough nature, and whether the firm is based in Shenzhen.

Given the previous findings that the effect of patents on firm income is statistically insignificant, we include both patent and R\&D variables in our heterogeneity analysis to ensure a comprehensive view. To this end, we augment our baseline regression model (Equation~\ref{equation 8}) by introducing interaction terms: 
\[
\text{Patent}_{i} \cdot \text{SubIndustry}, \quad \text{Patent}_{i} \cdot \text{Breakthrough}, \quad \text{Patent}_{i} \cdot \text{Base};
\]
\[
\text{R\&D}_{i} \cdot \text{SubIndustry}, \quad \text{R\&D}_{i} \cdot \text{Breakthrough}, \quad \text{R\&D}_{i} \cdot \text{Base}.
\]

The regression results are presented in Table~\ref{table: hetero}. For reference, the coding schemes used for sub-industries and firm base location are displayed in Tables~\ref{tab:industry_code_eng} and~\ref{tab:sz_base_eng}, respectively.

Regarding the technological breakthrough dimension, we adopt a composite quality score based on three indicators: (1) whether the project involves disruptive technologies, (2) whether the product or service can substitute for imported technologies, and (3) whether it addresses bottlenecks in upstream or downstream supply chains. A firm receives one point for satisfying each criterion, resulting in a patent quality score ranging from 0 to 3.

According to the results, we observe that \textit{R\&D input} has significantly greater marginal returns to income in several industries relative to the baseline category, Semiconductor and Integrated Circuit. In particular, R\&D is more productive in the Smart Terminal, Digital Creativity, and Ultra HD Video Display sectors. These findings suggest that R\&D investment translates more effectively into performance gains in industries with faster innovation cycles and higher output elasticity.

In addition, R\&D appears to have a stronger effect on income for firms headquartered in Shenzhen. This is consistent with prior literature highlighting Shenzhen’s role as a high-tech cluster \citep{gaoxinservjice3}, where supporting infrastructure, policy incentives, and talent pools amplify the returns to innovation inputs.

Taken together, these heterogeneity results demonstrate that the effectiveness of innovation is not uniform across firms. Instead, it is shaped by industry-specific dynamics and regional innovation ecosystems. Our findings highlight the importance of contextualizing innovation policies to account for structural variation in the translation of inputs into performance.

\begin{table}[htbp]\centering
\caption{Industry Code and English Description}
\label{tab:industry_code_eng}
\begin{tabular}{c l}
\hline
Code & Industry \\
\hline
1 & Semiconductor and Integrated Circuit \\
2 & Smart Terminal \\
3 & Network and Communication \\
4 & Software and Information Services \\
5 & Artificial Intelligence \\
6 & Digital Creativity \\
7 & Smart Sensor \\
8 & Ultra HD Video Display \\
9 & Future Industry / Optical Information \\
\hline
\end{tabular}
\end{table}

\begin{table}[htbp]\centering
\caption{Whether Company is Based in Shenzhen}
\label{tab:sz_base_eng}
\begin{tabular}{c l}
\hline
Code & Description \\
\hline
0 & Based in Shenzhen \\
1 & Not Based in Shenzhen \\
\hline
\end{tabular}
\end{table}

\begin{table}[htbp]\centering
\vspace{-3cm}
\caption{Heterogeneity tests}
\label{table: hetero}
\hspace*{-1.2cm}
\begin{tabular}{lcccccc} \hline
 & (1) & (2) & (3) & (4) & (5) & (6) \\
VARIABLES & Income & Income & Income & Income & Income & Income \\ \hline
 &  &  &  &  &  &  \\
2.industry\_Patents & 0.7948 &  &  &  &  &  \\
 & (0.6769) &  &  &  &  &  \\
3.industry\_Patents & -1.2380 &  &  &  &  &  \\
 & (1.0439) &  &  &  &  &  \\
4.industry\_Patents & -0.1259 &  &  &  &  &  \\
 & (0.3857) &  &  &  &  &  \\
5.industry\_Patents & 0.5184 &  &  &  &  &  \\
 & (0.5914) &  &  &  &  &  \\
6.industry\_Patents & -0.1399 &  &  &  &  &  \\
 & (0.3849) &  &  &  &  &  \\
7.industry\_Patents & 0.2840 &  &  &  &  &  \\
 & (0.4234) &  &  &  &  &  \\
8.industry\_Patents & -3.3002 &  &  &  &  &  \\
 & (4.6701) &  &  &  &  &  \\
9.industry\_Patents & -0.4097 &  &  &  &  &  \\
 & (0.9788) &  &  &  &  &  \\

2.industry\_R\&D &  & 0.7844*** &  &  &  &  \\
 &  & (0.2693) &  &  &  &  \\
3.industry\_R\&D &  & -0.5239 &  &  &  &  \\
 &  & (0.3950) &  &  &  &  \\
4.industry\_R\&D &  & -0.0308 &  &  &  &  \\
 &  & (0.1441) &  &  &  &  \\
5.industry\_R\&D &  & 0.2205 &  &  &  &  \\
 &  & (0.2398) &  &  &  &  \\
6.industry\_R\&D &  & 0.6021** &  &  &  &  \\
 &  & (0.2644) &  &  &  &  \\
7.industry\_R\&D &  & -0.1118 &  &  &  &  \\
 &  & (0.2188) &  &  &  &  \\
8.industry\_R\&D &  & 0.4532*** &  &  &  &  \\
 &  & (0.1427) &  &  &  &  \\
9.industry\_R\&D &  & -0.2080 &  &  &  &  \\
 &  & (0.1850) &  &  &  &  \\

1.Quality\_Patent &  &  & 0.4813 &  &  &  \\
 &  &  & (0.3499) &  &  &  \\
2.Quality\_Patent &  &  & 0.8125* &  &  &  \\
 &  &  & (0.4722) &  &  &  \\
3.Quality\_Patent &  &  & 0.6361 &  &  &  \\
 &  &  & (0.4172) &  &  &  \\

1.Quality\_R\&D &  &  &  & 0.0605 &  &  \\
 &  &  &  & (0.1877) &  &  \\
2.Quality\_R\&D &  &  &  & -0.0836 &  &  \\
 &  &  &  & (0.1705) &  &  \\
3.Quality\_R\&D &  &  &  & -0.1017 &  &  \\
 &  &  &  & (0.1999) &  &  \\

1.SZBase\_patent &  &  &  &  & -0.1730 &  \\
 &  &  &  &  & (0.1592) &  \\

1.SZBase\_R\&D &  &  &  &  &  & 0.2764** \\
 &  &  &  &  &  & (0.1274) \\

\end{tabular}
\end{table}

\section{Conclusion}

This study examines the relationship between innovation activities—particularly patents and R\&D investment—and firm-level productivity among early-stage high-tech firms in China. Our analysis yields several key findings.

First, contrary to conventional expectations, patent output does not significantly affect firm income or profit. While we carefully addressed potential endogeneity and removed confounding effects such as signaling, the insignificance of patents suggests that they may function more as symbolic or nominal outputs rather than as effective drivers of firm performance. This paradox leads us to reflect on the actual role of patents in early-stage innovation, particularly in emerging markets where the institutional value of patents may diverge from their economic substance.

In contrast, R\&D investment shows a robust and significant positive effect on firm productivity across multiple specifications. Mechanism analysis reveals that R\&D contributes not only through potential technological development, but also via improving firm environments and enhancing policy eligibility. However, we find little evidence that R\&D reshapes employee structures, which suggests that its impact is more systemic than talent-driven. Moreover, the R\&D intensity is significantly influenced by firm valuation and investment inflow, forming a feedback loop between capital and innovation.

Further, heterogeneity analysis highlights that the productivity effect of R\&D varies by sub-industry and firm location. Sectors like smart terminals and digital creativity exhibit higher R\&D returns, likely due to more efficient innovation-commercialization pathways. Firms based in Shenzhen also display stronger R\&D effectiveness, consistent with the advantages of policy support and local innovation ecosystems.

Taken together, these findings suggest that for early-stage firms, substantive innovation efforts—reflected through capital-backed R\&D—are more indicative of productivity gains than the mere accumulation of patents. From a policy perspective, this implies the need to shift evaluation criteria away from patent quantity toward meaningful innovation processes. Supporting firms through targeted, R\&D-linked incentives, particularly in high-return industries and innovation clusters, may better promote sustainable technological growth.

\medskip
\noindent
\textbf{Limitations and Future Directions.} 

While our study contributes to a growing understanding of innovation dynamics in early-stage high-tech firms, it is important to acknowledge several limitations—many of which are rooted in the inherent challenges of working with startup data.

First, a key constraint arises from data availability. High-tech startups often refrain from public disclosure, resulting in sample selection bias within academic literature, which tends to focus on more mature firms. This opacity makes it particularly difficult to access granular, longitudinal data required for comprehensive analysis. Despite this challenge, we have endeavored to make the most of the available information and believe our findings offer meaningful insights under these constraints.

Second, our dataset lacks multi-year, firm-level data on employee structures. Although our mechanism analysis regarding human capital attraction suggested limited evidence for this channel, the absence of detailed temporal hiring data introduces a potential bias that we cannot fully rule out. Additionally, the limited information on annual shifts in employee composition prevents us from conducting more precise tests of organizational transformation as a result of R\&D investment.

In sum, while our empirical strategy is well-grounded given current data conditions, we call upon future researchers to explore similar questions using richer and more granular datasets. Enhanced data on hiring timelines, R\&D allocation breakdowns, and internal organizational dynamics would allow for a more definitive evaluation of the proposed mechanisms and strengthen the empirical foundation for innovation policy targeting early-stage firms.

\newpage
\bibliographystyle{apalike}
\bibliography{aguiar,pericles,pericles_1,pericles_2,S0923474800000254,pericles_1097026626,S0305750X18303954,S0167718707001373,S0048733311000680}
\end{document}